\documentstyle[psfig,epsfig,graphicx]{mn}

\newcommand{\be}{\begin{equation}}
\newcommand{\ee}{\end{equation}}
\newcommand{\ba}{\begin{eqnarray}}
\newcommand{\ea}{\end{eqnarray}}

\newcommand{\nn}{\nonumber \\}

\newcommand{\n}{\mbox{\boldmath $n$}}

\newcommand{\mnras}{MNRAS}

\def\gs{\mathrel{\raise1.16pt\hbox{$>$}\kern-7.0pt %
\lower3.06pt\hbox{{$\scriptstyle \sim$}}}}         %
\def\ls{\mathrel{\raise1.16pt\hbox{$<$}\kern-7.0pt %
\lower3.06pt\hbox{{$\scriptstyle \sim$}}}}         %

\title[On Scale-Dependent Cosmic Shear Systematic Effects]{On Scale-Dependent Cosmic Shear Systematic Effects}

\author[Kitching et al.]
       {T. D. Kitching$^1$\thanks{t.kitching@ucl.ac.uk}, A. N. Taylor$^2$, M. Cropper$^{1}$, H. Hoekstra$^{3}$, 
         \newauthor
         R. K. E. Hood$^{1}$, R. Massey$^{4,5}$, S. Niemi$^{1}$\\
         $^1$Mullard Space Science Laboratory, University College London, Holmbury St Mary, Dorking, Surrey RH5 6NT, UK\\
         $^2$SUPA, Institute for Astronomy, University of Edinburgh, Royal Observatory, Blackford Hill, Edinburgh, EH9 3HJ, UK\\
         $^3$Leiden Observatory, Leiden University, PO Box 9513, 2300 RA, Leiden, the Netherlands\\
         $^4$Institute for Computational Cosmology, Durham University, South Road, Durham DH1 3LE, UK\\
         $^5$Centre for Extragalactic Astronomy, Durham University, South Road, Durham DH1 3LE, UK}

\date{}

\pagerange{\pageref{firstpage}--\pageref{lastpage}}

\pubyear{2015}

\begin{document}

\maketitle

\label{firstpage}

\begin{abstract}
In this paper we investigate the impact that realistic scale-dependence systematic effects may have 
on cosmic shear tomography. We model spatially varying residual ellipticity and size variations in weak lensing 
measurements and propagate these through to predicted changes in the uncertainty and bias of cosmological parameters. 
We show that the survey strategy -- whether it is regular or randomised -- is an important factor in 
determining the impact of a systematic effect: a purely randomised survey strategy produces the smallest biases, at the 
expense of larger parameter uncertainties, and a very regularised survey strategy 
produces large biases, but unaffected uncertainties. However,  
by removing, or modelling, the affected scales ($\ell$-modes) in the regular cases the biases are reduced to negligible levels. 
We find that the integral of the systematic power spectrum is 
not a good metric for dark energy performance, and we advocate that systematic effects should be modelled accurately 
in real space, where they enter the measurement process, and their effect subsequently propagated 
into power spectrum contributions. 
\end{abstract}

\begin{keywords}
Cosmology: theory -- large--scale structure of Universe
\end{keywords}

\section{Introduction}
\label{Introduction}
Weak gravitational lensing is the effect whereby the image of a background object is 
distorted as a result of the intervening mass, causing tidal effects along the line of sight. 
An effect of weak lensing on the images of galaxies is to cause the third eccentricity (or third 
flattening) - colloquially referred to as `ellipticity' - of the images to change. This change, to first order, 
in ellipticity is known as `shear'. Galaxy images can be affected by shear distortions 
as a result of the gravitational potential around galaxies or clusters. They can also be 
affected by lensing from all scales in the cosmic web (large-scale structure) so that every galaxy is 
sheared by a very small amount; this is known as cosmic shear. 

Cosmic shear is a particularly interesting phenomenon because the amplitude of the excess probability
for any pair of galaxies to have aligned shear distortions -- the correlation function -- is related
to the power spectrum of matter density perturbations, the growth of structure, and the
distance-redshift relation. In this paper, and in the majority of the theoretical and methodological
literature on cosmic shear the harmonic (Fourier) transform of the correlation function is used,
which is known as the power spectrum. This statistic depends on several aspects of the
cosmos that have a strong sensitivity to variations in cosmological parameters, and as a function of redshift.
In particular cosmic shear is a statistic that is sensitive to changes in the dark energy equation of state.

Supplementing cosmic shear information 
with redshift information of galaxies is known as 3D cosmic shear (see Kitching et al., 2014 and 
references therein for an exposition). 
Cosmic shear `tomography' is an approximation of 
3D cosmic shear by assuming a fixed linear relation between the radial and azimuthal wavenumbers 
for the 3D shear field (the Limber approximation), and a binning in 
redshift (Kitching, Heavens, Miller, 2011). 
For a reviews of weak lensing and cosmic shear see 
for example Bartelmann \& Schneider (2001), Hoekstra \& Jain (2008) and Kilbinger (2014). 
In this paper we will focus on the requirements on systematic effects set when using  
cosmic shear tomography for cosmology.  

Because cosmic shear can potentially measure the dark energy equation of state very accurately, there are 
several experiments either on-going, or planned, that will use cosmic 
shear. From the ground these are KiDS\footnote{{\tt http://kids.strw.leidenuniv.nl}}, 
DES\footnote{\tt www.darkenergysurvey.org}, HSC\footnote{\tt www.subarutelescope.org/Projects/HSC/}, LSST\footnote{\tt http://www.lsst.org}, and 
from space \emph{Euclid}\footnote{{\tt http://euclid-ec.org}} and 
\emph{WFIRST-AFTA}\footnote{{\tt http://wfirst.gsfc.nasa.gov}}. 

The shear effect is a very small change in the ellipticity of a galaxy image: a change 
in the third flattening (`ellipticity') of approximately $10^{-2}$ (or, $\sim 10\%$ of the amplitude, 
which is approximately $10^{-1}$ on average). The 
measurement from data is complicated 
in several respects: even without the shear effect galaxy images would appear elliptical, there are  
`intrinsic' ellipticity correlations (see Troxel \& Ishak, 2014, Joachimi et al., 2015 for reviews); the measurement of ellipticity is 
affected by noise (Viola, Kitching, Joachimi, 2014) and modelling uncertainities (if a galaxy 
model is used, see for example Kitching et al., 2012; Voigt \& Bridle, 2010); telescopes have an impulse-response, 
or point-spread-function 
(PSF) that blurs images (and, for ground-based telescopes, atmospheric effects further blur the images), 
and the PSF is typically also elliptical to some degree; and CCD detectors 
themselves can also degrade ellipticity as a result of radiation damage (CTI), manufacturing errors, or 
properties inherent to the devices. Therefore the control of systematic effects in experiments that 
wish to measure cosmic shear need to be characterised and accounted for in a rigorous manner. 

There have been a number of studies of systematic effects that can degrade cosmic shear measurements (for example 
Massey et al., 2013; Kitching et al., 2009; Amara \& Refregier, 2008; and references therein). 
These have typically looked at requirements either on an individual galaxy basis, or under the assumption 
that systematic effects are not dependent on angular scale (or a random dependency). Some exceptions include the 
galaxy ellipticity measurement simulations in Kitching et al. (2012) that investigated the change in power 
spectra from imperfect measurements, and Hoekstra (2004) who compared PSF modelling errors to the cosmic shear signal in data. 
Those studies that have included some 
scale-dependence have made an assumption that all systematic effects have 
scale-dependent functional behaviour that mimics a cosmological signal, the requirement 
used being an integral over all scales used in the analysis. This was the approach taken in 
Massey et al. (2013) and Cropper et al. (2013) that propagated requirements on systematic effects 
through to the performance on cosmological parameters. 

In this paper we look at more realistic scenarios for the scale-dependence of systematic effects, on 
CCD detector, field-of-view, and larger scales. To realise a concrete version of 
these scales, we assume the baseline design of the Euclid hardware (Laureijs et al., 2011).
We then propagate the expected angular behavior of systematic effects 
through to cosmic shear tomography cosmological parameter predictions. We also present a simple way to mitigate 
their impact by removing the scales at which the systematic has an effect.  

The paper is presented as follows. In Section \ref{Method} we outline the basic methodology. In Section \ref{PS} we present 
some simple examples to demonstrate the main conclusions of the paper. In Section \ref{Results} we present some 
more realistic cases, and an application to charge transfer inefficiency requirements. In Section \ref{Conclusion} we present some
conclusions. 

\section{Method}
\label{Method}
The method we use is the following: we introduce realistic spatially varying systematics into our images, 
then measure the power spectra of these, 
and propagate them into the shear power spectra. We use the notation from Massey et al. (2013) and Cropper et al. (2013) 
where the observed tomographic cosmic shear power spectra $C^{\rm obs}_{ij,\ell}$ can be related to the 
true (systematic free) power spectra $C^T_{ij,\ell}$ given by 
\be 
\label{cce1}
C^{\rm obs}_{ij,\ell}=(1+{\mathcal M}_{ij,\ell})C^T_{ij,\ell}+{\mathcal A}_{ij,\ell}
\ee 
where $\ell$ labels the angular wavenumber, and $ij$ labels a redshift bin pair for which $i=[1, N_{\rm bin}]$. We 
consider in this expression a `multiplicative' term ${\mathcal M}_{ij,\ell}$ and an `additive' term 
${\mathcal A}_{ij,\ell}$; here we label the redshift bin pair $ij$ for generality, but for most of this paper 
we assume that they are redshift-independent effects and so do not use these labels. Equation (\ref{cce1}) 
neglects mode-mixing that could contaminate $E$ and $B$-mode power,  
intrinsic alignment terms (as shown in Kitching et al., 2012, Appendix A),  and assumes that the multiplicative term is 
uncorrelated in $\ell$ (this term should be $\sum_{\ell'}{\mathcal M}_{\ell\ell'}C_{ij,\ell'}$, so we are assuming no 
off-diagonal terms in ${\mathcal M}_{\ell\ell'}$, we will investigate this further in future work). 
A `systematic' power spectrum can then be defined as 
\be
C^{\rm sys}_{ij,\ell}={\mathcal M}_{\ell}C^T_{ij,\ell}+{\mathcal A}_{\ell}.
\ee
The shear power spectrum can be related to the matter power spectrum $P(k; r(z))$, 
where $k$ $h$Mpc$^{-1}$ is a radial wavenumber and 
$r(z)$ is a comoving distance, and angular diameter distance through 
\be 
C^T_{ij,\ell}=A\int_0^{r(z_H)}{\rm d}r' \frac{W_i(r')W_j(r')}{a(r')^2}P(\ell/r'; r[z])
\ee
where $A=(3\Omega_{\rm M}H_0^2/2c^2)^2$ 
and $\Omega_{\rm M}$ is the dimensionless matter density, $H_0$ is the current value of the Hubble 
parameter, $z_H$ is the redshift 
of the cosmic horizon, $c$ is the speed of light in a vacuum, and $a(r)$ is the dimensionless scale factor. 
The weight function is $W_i(r)=\int_r^{r(z_H)}{\rm d}{r'} p(r'|r[z_i])f_K(r'-r)/f_K(r')$, with $i$ labelling a 
redshift bin,  where 
$f_K(r)=\sinh(r)$, $r$, $\sin(r)$ for curvatures of 
$K=-1$, $0$, $1$, and $p(r'|r)$ is the probability that a galaxy with comoving distance $r$ is observed at distance $r'$. This 
representation of the shear power spectrum assumes the Limber approximation, a spherical 
Bessel transform to comoving distance, and a 
binning in redshift. For a derivation of this from the full 3D cosmic shear power spectrum see Kitching, Heavens, Miller (2011). 

\subsection{Systematic Propagation}
In Massey et al. (2013) and Cropper et al. (2013) the multiplicative and additive terms were linked to systematic changes in 
convolutive effects on the image (i.e. PSF effects and similar), non-convolutive effects (i.e. CTI effects and similar) and inaccuracies in the measurement of 
ellipticities. These papers looked at correlations of complex residuals $\langle \delta x \delta x^*\rangle$ where 
$\delta x=x_M-x_T$, the difference between the measured and true values of some quantity which is then 
decomposed into a bias and variance $\langle \delta x \delta x^*\rangle=\langle (x_M-\overline{x_M})^2\rangle
+ \langle (\overline{x_M} - x_T)^2\rangle=\sigma^2(x_M)+{\rm b}^2(x_M)$, where an bar denotes a mean quantity. 
Requirements can then be set on the bias and the variance that are functions of $x_M$. 
In this paper we expand the variance to the Fourier power spectrum of the variable, and treat the bias as a constant 
term. Massey et al. (2013) used an unconventional angle bracket notation when 
referring to the power spectra, as explained in Cropper et al. (2013).

In this paper we are concerned with systematic effects that leave a residual, unknown pattern in the data that may occur after 
modelling and correction. We will focus on the residuals that may occur as a result of PSF measurement, CTI correction and 
shape measurement such that the variables that we model are: residual PSF ellipticity $\epsilon_{\rm PSF}$; residual 
PSF size measurements 
$R^2_{\rm PSF}$; residual CTI ellipticity $\epsilon_{\rm CTI}$; residual CTI size measurements $R^2_{\rm CTI}$; and 
systematic effects 
in shape measurement, $\mu$ and $\alpha$, that contribute to the multiplicative and additive parts of the power spectrum 
respectively. Throughout we will refer to normalised sizes so that all quantities are divided 
by their mean values e.g. $R^2_{\rm PSF}=R^2_{\rm PSF, unnormalised}/\overline{R^2}_{\rm PSF, unnormalised}$. 
These quantities are related to the terms ${\mathcal A}$ and ${\mathcal M}$ in Massey et al. (2013), and are summarised 
in Cropper et al. (2013), which we reproduce in the following equations using the power spectrum notation. 
The multiplicative and additive terms have weighted contributions from the residual systematic effects given by  
\ba
\label{Mwieght}
{\mathcal M}_{\ell}
&=&m_1 \overline{R^2_{\rm PSF}}+m_2\overline{R_{\rm CTI}}\nn
&+&m_3 R^{-4}_{\rm PSF} C^{R^2_{\rm PSF}}_{\ell}+m_4 R^{-2}_{\rm CTI} C^{R_{\rm CTI}}_{\ell}\nn
&+&m_5 \overline{\delta\mu},
\ea
and
\ba
\label{Awieght}
{\mathcal A}_{\ell}
&=&a_1 C^{\epsilon_{\rm PSF}}_{\ell}+a_2 C^{\epsilon_{\rm CTI}}_{\ell}\nn
&+&a_3 R^{-4}_{\rm PSF} C^{R^2_{\rm PSF}}_{\ell}+a_4 R^{-2}_{\rm CTI} C^{R_{\rm CTI}}_{\ell}\nn
&+&a_5 C^{\alpha}_{\ell}, 
\ea
where overline denotes the mean of a quantity. In converting these quantities 
from the real-space notation to power spectra we note the following aspects. 
For constants (like the $m_1$ and $m_2$ terms), the Fourier transform results in a delta function, but the propagation through to 
multiplicative bias also involves a convolution. Carrying out the convolution turns them into constant multiplicative biases in 
the power spectra. 
The remaining terms, which still convolve $C_{\ell}$, are then treated as diagonal, and hence $\ell$-dependent multiplicative biases.
The terms $m_3$ and $m_4$ in Cropper et al. (2013) only refer to errors, but as stated in Massey et al. (2013) they 
``[Ignore] a bias term in ${\mathcal M}$ proportional to the square of one already present (therefore negligible if the bias is small)'', 
so these convert to $C_{\ell}$ terms, despite no bias being refered to.

The constants $a_i$ and $m_i$ can be derived from 
galaxy and instrumental properties, and have values, taken from Cropper et al. (2013), of: 
$a_1=0.1$, $a_2=0.74$, $a_3=0.001$, $a_4=0.0042$, $a_5=0.001$, 
and $m_1=1.20$, $m_2=0.34$, $m_3=0.18$, 
$m_4=0.015$, $m_5=1.20$. 
We note that there are two errata in Cropper et al. (2013) 
for the numerical values of the terms: $a_4'=4a_3'$ should be $0.004$ not $0.0042$, and $m_4'=(m_2')^2/4$ should be $0.007225$ 
not $0.0075$. These slightly increased values add some flexibility  
to the requirements by inceasing the contribution from these systematic effects to the overall systematic power spectrum,
so we use the larger values in this paper.
These expressions make the same assumptions as in Massey et al. (2013), that there is no  
correlation between systematics, and that some quantities are 
negligible with respect to others (for example the 
potential mode-mixing from correlations in $\mu$). 

We also note that the 
$m_5$ term in Cropper et al. (2013) could have read 
$m_5(\langle\delta\mu\rangle+(1/2)[\langle\delta\mu\rangle^2+\sigma^2(\mu)])$ (propagating both terms from Cropper et al., 2013, 
equation 4, rather than only the $\langle m\rangle$ term). However as the $\sigma(\mu)$ term is assumed to be negligble (calibration emulations of data 
can be increased until this is the case), to linear order the term was used correctly and this is what 
we use in this paper. We have generalised the 
assessment of systematics to include scale-dependence, 
and we leave the investigation of these assumptions to future work. 

\subsection{Procedure}
\label{Procedure}
Given the equations (\ref{Awieght}) and (\ref{Mwieght}) we can now assess the impact of spatially varying, scale-dependent, effects
on cosmic shear tomographic power spectra. The procedure that we follow is the following:
\begin{enumerate}
\item
We create 2D residual ellipticity and size fields for each of the systematic effects. These can be any form, but in this
paper we use prescriptions that mimic realistic scenarios.
\item
We compute the power spectrum for each field. This uses the steps shown in Kitching et al. (2012) 
that take the Fourier transform of the
field, rotate in Fourier space to an E/B mode frame, and take the average of the Fourier transform of the real part 
in shells in angular wavenumber $\ell$.
\item
We combine these power spectra using equations (\ref{Awieght}) and (\ref{Mwieght}) to 
create additive ${\mathcal A}$ and
multiplicative ${\mathcal M}$ functions.
\item
We propagate the systematic power spectrum into Fisher matrix predictions to assess the change in 
cosmological parameter error (uncertainty), and bias, caused by the systematic.
\end{enumerate}

\subsection{Prediction Method}
\label{PM}
To compute expected cosmological parameter errors we use the Fisher matrix formalism presented in 
Hu (1999) for cosmic shear tomography. 
This results in a matrix $F_{\alpha\beta}$ (the Greek letters denote cosmological parameter pairs) where the
$[(F^{-1})_{\alpha\alpha}]^{1/2}$ is a vector of expected, marginalised, cosmological parameter error.
To compute the expected biases in cosmology parameters we use the formalism described in Kitching et al. (2009) where the predicted shift
in a parameter $\alpha$ caused by a systematic effect is $b_{\alpha}=-(F^{-1})_{\alpha\beta}B_{\beta}$, where the vector $B$ for each 
parameter $\beta$ is 
$B_{\beta}=\sum_{ij,\ell}(1/\sigma_C^2)C^{\rm sys}_{ij,\ell}(\partial C_{ij,\ell}/\partial \beta)$ and the error on the power spectra $\sigma_C$ is given in Hu (1999).

We use a CDM cosmology with a varying dark energy equation of state, where the free parameters are
$\Omega_{\rm M}$, $\Omega_{\rm B}$, $\sigma_8$, $w_0$, $w_a$, $h$, $n_s$ (respectively the dimensionless matter density; dimensionless
baryon density; the amplitude of matter fluctuations on $8$Mpc scales -- a normalisation of the power spectrum of matter perturbation; 
the dark energy equation of state parameterised by $w(z)=w_0+w_a z/(1+z)$; the Hubble parameter $h=H_0/100$kms$^{-1}$Mpc$^{-1}$; and
the scalar spectral index of initial matter perturbations). For each parameter we use the \emph{Planck} maximum likelihood
values (Planck, Paper XVI, 2013) about which we take derivatives of the power spectra for the Fisher matrix and bias vector. All parameter
errors and biases we quote are marginalised over all other parameters in this set. We use the {\tt camb sources}\footnote{
{\tt http://camb.info/sources/}} code to compute the
cosmic shear tomographic power spectra\footnote{Matlab code to reproduce the results of this paper is available on request.}. We use a
maximum radial wavenumber of $k_{\rm max}=5h$Mpc$^{-1}$ and a corresponding redshift-dependent maximum $\ell$-mode of $\ell=k_{\rm max}r[z]$.

The cosmic shear survey we assume is a \emph{Euclid}-like experiment that has an area of
$15$,$000$ square degrees, a median redshift of
$z_m=0.9$, a number density of $30$ galaxies per 
square arcminute, with a number density distribution $n(z)$ given in Taylor et al. (2006), and 
a photometric redshift probability distribution that is assumed to be Gaussian with a standard deviation of $\sigma(z)=0.05(1+z)$. These
characteristics are described in Laureijs et al. (2011), and are the same as those used in Massey et al. (2013). We assume that a field-of-view
is $0.5$ square degrees, covered by a $6\times 6$ array of detectors, that are assumed to be four-side buttable (i.e. no gaps 
between the chips).

\section{An investigation of Power Spectrum Requirements}
\label{PS}
In Massey et al. (2013) requirements were set by Monte Carlo evaluation of the 
functional form of ${\mathcal A}_{\ell}$ and ${\mathcal M}_{\ell}$, and estimating the error, and bias, on the dark energy equation of state 
using the method described in Kitching et al. (2009). 
This found the ``worst case'' variation of these functions 
arising from the selection of a 
functional behaviour of ${\mathcal A}$ and ${\mathcal M}$ that most closely matched 
the derivative of the shear 
power spectrum with respect to the dark energy equation of state $\partial C_{ij,\ell}/\partial w_0$, 
with additional considerations arising because of the multiple tomographic bins 
and degeneracies with other cosmological parameters. Once this function was found the 
mean values of the ${\mathcal A}$ and ${\mathcal M}$ were computed and this was used
to set requirement on PSF, CTI and shape measurement quantities through the weighting scheme in equations (\ref{Mwieght}) and (\ref{Awieght}). The requirements on 
the mean integrated values\footnote{From Cropper et al., 2013, where these are labelled ${\mathcal A}'$ and ${\mathcal M}'$.} are:  
$\overline{{\mathcal A}}\leq 2.6\times 10^{-7}$ and $\overline{{\mathcal M}}\leq 1.4\times 10^{-2}$ using a maximum wavenumber of $\ell_{\rm max}=5000$ 
to avoid the highly non-linear regime.  A more stringent 
combined requirement on the integrated quantity\footnote{The integration limits were not defined for this quantity but in the text the range $10 < \ell < 2\times 10^4$ was specified.} $(1/2\pi)\int {\rm d} \ell (\ell+1) C^{\rm sys}_{ij,\ell}\leq 10^{-7}$ was 
set by Amara \& Refregier (2008), which they termed $\sigma_{\rm sys}^2$.  

This was a conservative approach to setting the requirement for systematic effects. 
In reality the propagation of systematic effects through to the dark energy equation of state depends on the amplitude, 
the total integral constraint of the systematic effects and the functional form 
in $\ell$ that the systematic effect causes.

\subsection{Simple Examples}
\label{simp}
\begin{figure*}
\centering
  \includegraphics[angle=90,clip=,width=2\columnwidth]{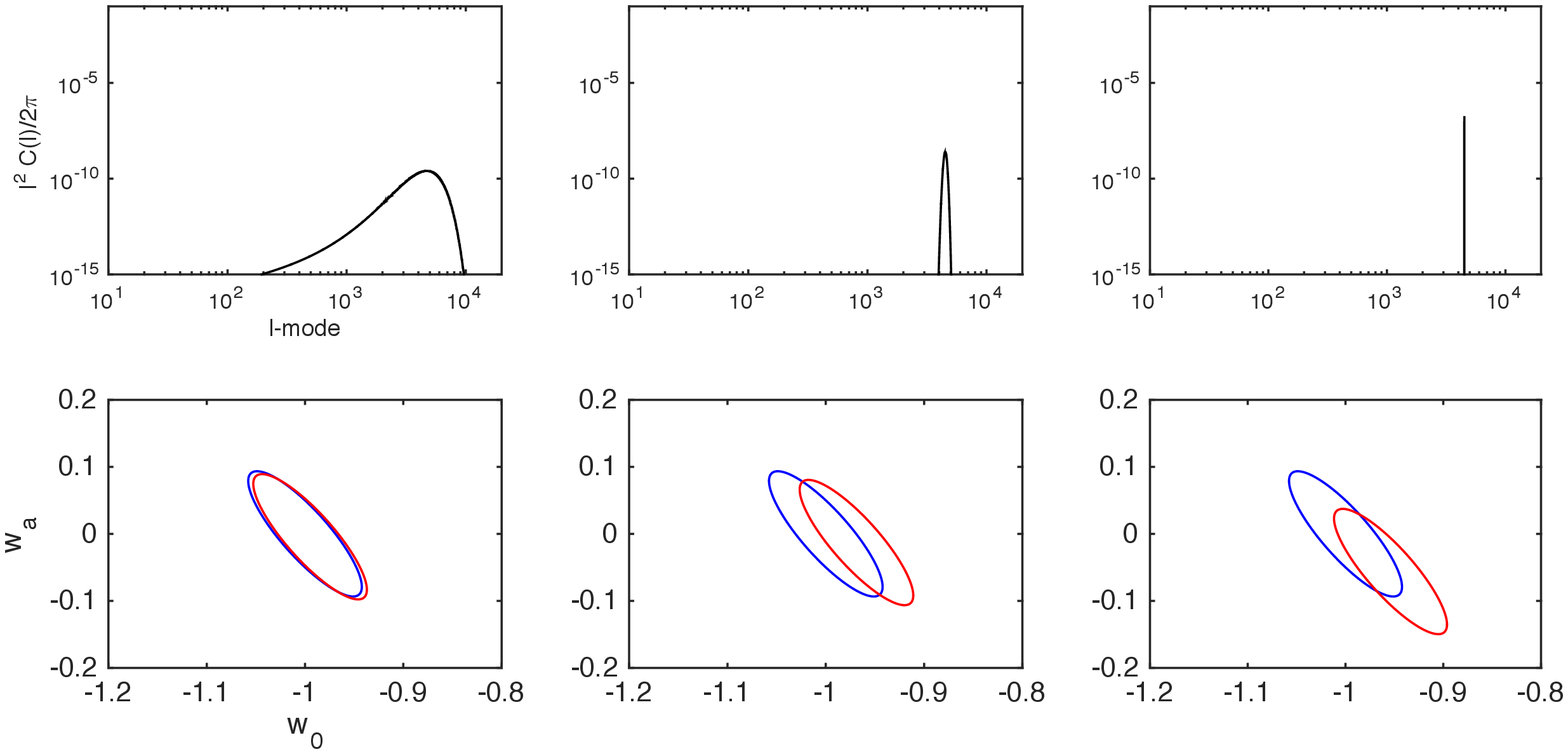}
 \caption{A simple example of the impact on the bias/error for the dark energy equation of state parameter $w_0$, 
for the survey parameters described in Section \ref{PM}. 
The systematic functional form in $\ell$ is set to a Gaussian in $\ell$ centered on $\ell=4500$ with three different 
full-width-half-maxima of $1000$, $100$ and $10$ from right to left panels. The upper panels show the 
systematic power spectra: the y-axes are $\ell^2 C(\ell)/(2\pi)$ and the x-axes are $\ell$-mode. 
The lower panels show the dark energy parameters with $w_a$ on the y-axes and $w_0$ on the x-axes; the contours are 
2-parameter 1-$\sigma$ predicted confidence ellipses; the blue shows the systematic-free case and the red shows the 
case with the systematic effect.}
 \label{test1}
\end{figure*}

In Figure \ref{test1} we show an example of an additive Gaussian systematic power spectrum in $\ell$, with a total integrated 
constraint of $\sigma_{\rm sys}^2=10^{-7}$ (using an $\ell_{\rm max}=20$,$000$). A wide range of biases meet this requirement. We find that a sharp high 
amplitude systematic effect in $\ell$ causes the largest bias, and a systematic with a slope of approximately unity in 
$\ell^2 C(\ell)$ has the smallest bias. 

\begin{figure*}
\centering
  \includegraphics[angle=0,clip=,width=2\columnwidth]{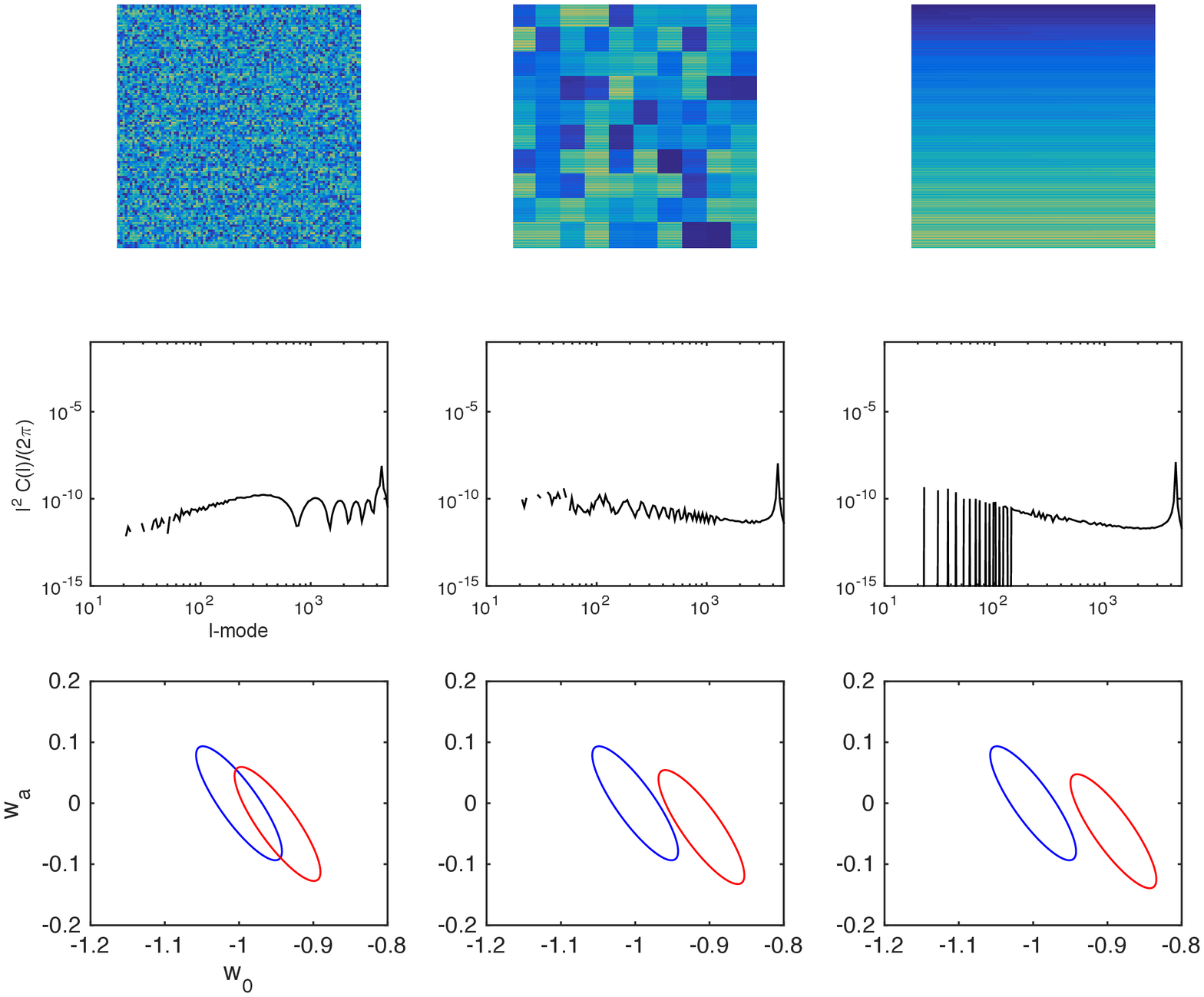}
 \caption{The upper panels show a simulated $50$ square degree patch with a residual ellipticity field whose amplitude is 
represent by the colour (yellow is maximum blue is minimum). The middle row of panels show the power spectra of the systematic effects all normalised 
to have an integrated value of $\sigma^2_{\rm sys}=10^{-7}$: the y-axes are $\ell^2 C(\ell)/(2\pi)$ and the x-axes are $\ell$-mode. 
The lower panels show the impact on the dark energy parameters with $w_a$ on the y-axes and $w_0$ on the x-axes; the contours are
2-parameter 1-$\sigma$ predicted confidence ellipses, the blue shows the systematic-free case and the red shows the
case with the systematic effect. The survey parameters are described in Section \ref{PM}. The left column is the case where 
the field-of-view patches are observed in a random order, as the CTI increases linearly over the observation sequence; the middle panels 
show the case where 5 square degree patches are observed contiguously before moving on to a new 5 square degree patch at random; the 
right panels show the case that the full 50 square degrees are observed in a regular rectilinear scan, with 
no randomness in the tiling, and an increasing ellipticity. The spike in power at high-$\ell$ comes from the CCD scale.}
 \label{test2}
\end{figure*}

In a second test, in Figure \ref{test2} we show an ellipticity field 
with different real-space 2D spatial patterns: from a rectilinear pattern to a randomised pattern. In this case we find that the 
regular pattern produces the largest bias, whereas the smallest bias is caused by the randomised pattern.

\begin{figure*}
\centering
  \includegraphics[angle=90,clip=,width=2\columnwidth]{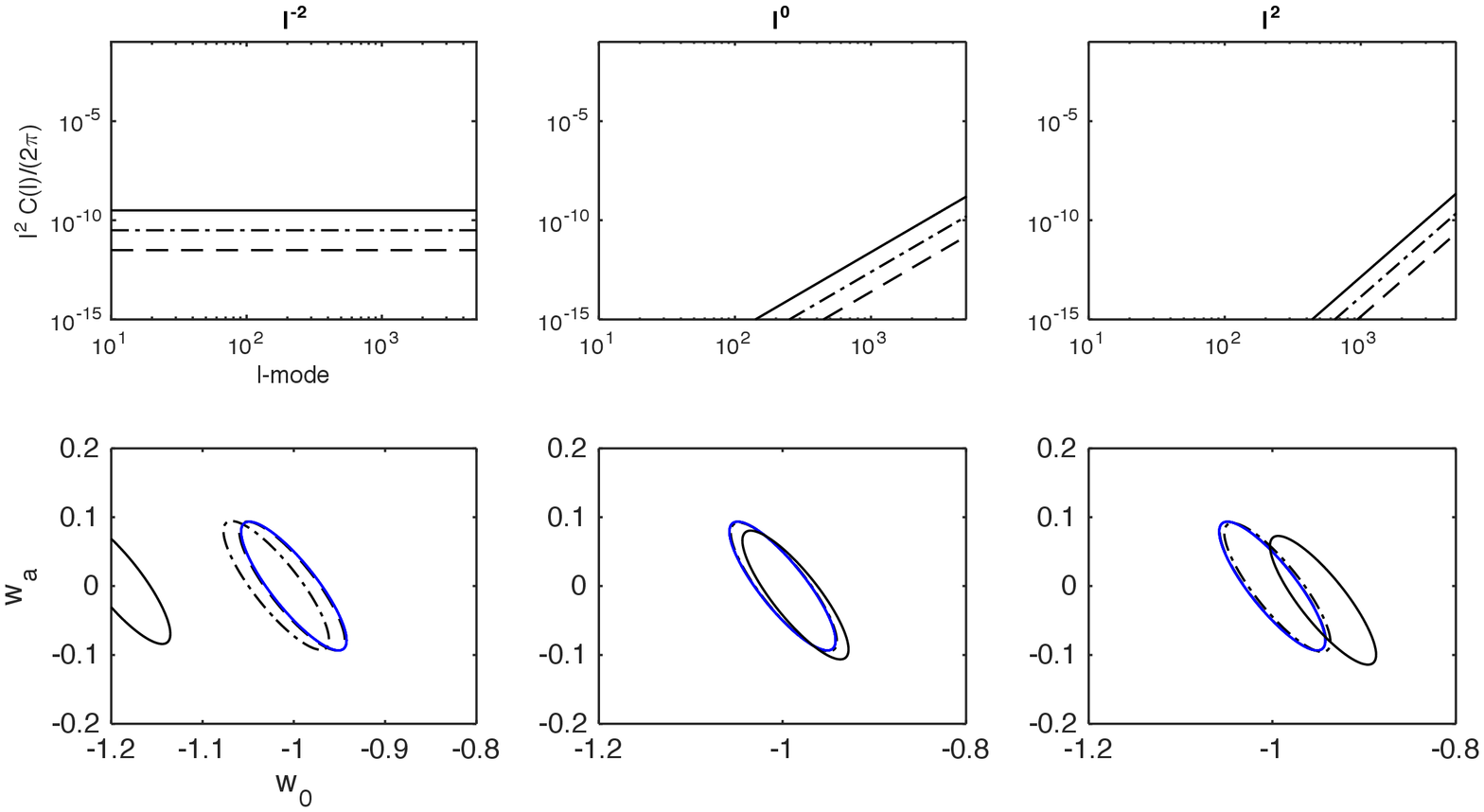}
 \caption{The upper panels show 
systematic power spectra, with a power-law functional form $\ell^n$ with slopes of $n=-2$, $0$ and $2$ as a function of $\ell$-mode: 
the y-axes are $\ell^2 C(\ell)/(2\pi)$ and the x-axes are $\ell$-mode. Three different integrated 
values of $\sigma^2_{\rm sys}$ are shown: $\sigma^2_{\rm sys}=10^{-7}$ (dashed), $\sigma^2_{\rm sys}=10^{-6}$ (dot-dashed) and $\sigma^2_{\rm sys}=10^{-5}$ (solid). 
The lower panels show the dark energy parameters with $w_a$ on the y-axes and $w_0$ on the x-axes; the contours are
2-parameter 1-$\sigma$ predicted confidence ellipses, the blue shows the systematic-free case and the black shows the
case with the systematic effect (dashed, dot-dashed and solid corresponding to the three cases in the upper panels).  
Where the contours used are not visible it is overlapping with the blue solid contour to within the thickness of the plotting line. 
The survey parameters are described in Section \ref{PM}.}
 \label{test3}
\end{figure*}

In fact a systematic effect that has the same functional form as pure shot noise will, by definition, not cause any biases 
in the inferred cosmological parameters, but the error bar on those parameters will increase. A shot noise 
power spectrum is flat in $\ell$-space, having equal power at all scales. In this case the 
systematic would add to the measured ellipticity shot noise, but we note that an unknown shot-noise-like component would still cause 
biases in the infered amplitude of the cosmic shear power spectra.
We numerically test this in Figure \ref{test3} where we show that the bias is minimised when an additive systematic effect 
has a slope of zero i.e. is consistent with shot noise. This conclusion is also supported by the Figures \ref{test1} and 
\ref{test2}. 

\subsection{Functional Sampling}
\begin{figure*}
\centering
  \includegraphics[angle=0,clip=,width=1\columnwidth]{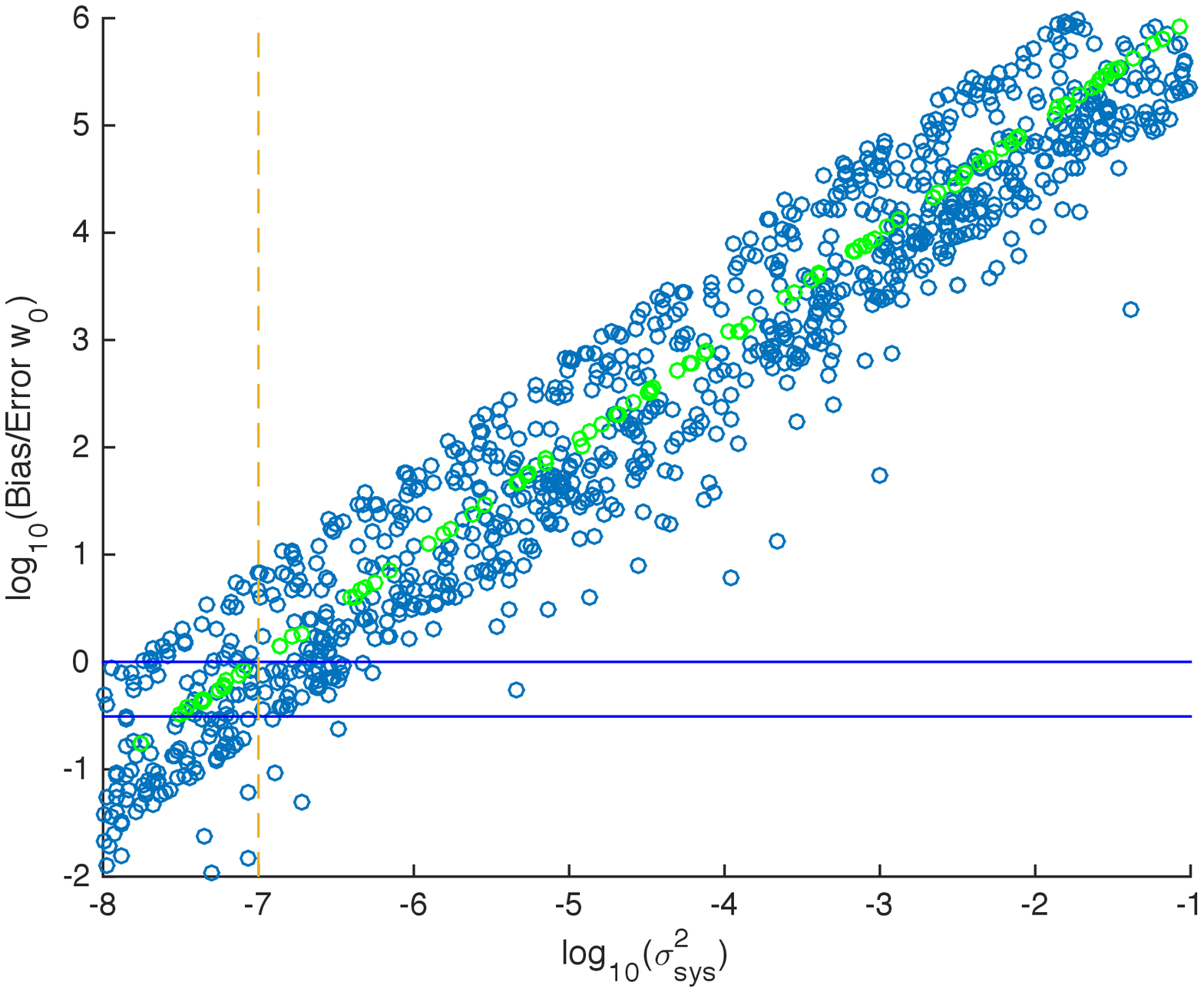}
 \caption{This Figure shows the impact that different functional forms for the 
systematic power spectrum have on the ratio of bias/error for the dark energy parameter $w_0$. The x-axes shows the 
integrated value of the systematic power spectrum $\sigma^2_{\rm sys}$, the y-axes shows the bias/error. The 
solid horizontal lines show bias/error$=1$ and $0.31$, the vertical (orange) dashed line shows $\sigma^2_{\rm sys}=10^{-7}$. 
The blue dots show the range of biases found when taking $1000$ realisations of the Guassian functional form described in 
Section \ref{simp}. The green dots show the upper limit of the biases caused by realisations of a binned functional form in $\ell$-mode 
(repeating the analysis of Massey et al., 2013). The survey parameters are described in Section \ref{PM}.}
 \label{redoMHK}
\end{figure*}

In these tests we also find that the total integral constraint on the systematic power spectrum is in fact not a good metric. 
A single integral constraint can allow for dramatically different functional behaviour in the power spectra, 
and eventual biases: we can find example for which almost no cosmological parameter biases are introduced, but for which requirements of 
are not met using the published limits on this integral. This is because 
in previous studies functional forms were chosen by randomly sampling from truncated functional expansions. 
For example $200$ bins 
in $\ell$ were used 
in Massey et al. (2014) and this is very unlikely to create an isolated spike in the reproduced power spectrum (a less 
than $1:10^{100}$ chance of producing a single $\ell$-mode peak at a height of $90\%$ of the maximum). 
We show this in Figure \ref{redoMHK} where we do the same test as in Massey et al. (2014), but also supplement this 
with realisations that randomly sample from a Gaussian with a width chosen between $\sigma(\ell)=[0, 500]$ and a mean chosen 
between $[100, 5000]$. We find that indeed with a truncated functional expansion an integrated limit of 
$\sigma_{\rm sys}^2\leq 10^{-7}$ does ensure that dark energy biases are below bias/error$=1$ (or the revised requirement of $0.31$ 
used in Massey et al., 2014), but that a more thorough 
functional search, including power spectra with spike-like functional features can produce biases that are an order of magnitude 
larger. 

These larger bias are caused because by spike-like features because it is not in fact a worst case when whole functional form of 
$C^{\rm sys}_{\ell}$ resembles the sensitivity of the power spectrum with respect to a parameter, but instead a sufficiently large 
change in amplitude at a single $\ell$-mode can cause a large bias. As an example consider fitting a straight line to some data  
points, one of which is offset by a very large amount: this would cause a bias in the fit of a gradient or offset despite only 
affecting a single data point (i.e. the systematic would not resemble the derivative of the model with resepct to either the 
gradient or offset at all data points). We note that the goodness of fit would be 
less in such a scenario, and indeed this could be measured 
to test if systematics are present, we leave this sophistication for future work.

We therefore conclude that integrated power spectrum requirements are not a good metric for dark energy performance.  
In Figure \ref{test2} we showed that the same integrated requirement can produce very different biases, 
and in Figure \ref{redoMHK} we showed that sampling parameter values of a different 
functional form can result in dramatically different dark energy biases. Consequently we advocate here an approach where the systematic ellipticity field 
in real space is modelled, and the effect on the predicted cosmological inference propagated through a 
power spectrum estimation of that 
field. In this paper we look at realistic angular behaviour of systematics to investigate realistic cases, 
and assess at what level there is margin\footnote{A systems engineering term that means the difference between the 
requirement assumed during development/construction and the true/applicable requirement, a difference that typically 
leads to increased actual performance of an experiment over its design expectation.} in the requirements for cosmic shear tomography. 

\section{Realistic Scenarios}
\label{Results}
To investigate the impact that spatially varying systematic effects may have we begin by defining three realistic scenarios. 
In each case we model a representative $50$ square degree patch of data, and assume that this pattern is repeated across whole 
survey area. These $50$ square degrees are tiled with $10$ arcsecond pixels which is the smallest angular scale that can be used in a cosmic shear 
analysis; this angular scale corresponds to radial scales ($k$-modes) of a few tens of Mpc$^{-1}$ at the highest redshifts in a typical weak lensing survey 
(see Kitching \& Taylor, 2010 for a discussion of this point). 
This is the same method that was used 
for Figure \ref{test2}. Throughout we use a flat-sky approximation in the power spectrum analyses.

\subsection{Experimental Setup}
The three scenarios are meant to represent some extreme cases where uncertainty in the PSF and CTI modelling 
result in  residuals that would formally not meet the requirements specified in Cropper et al. (2013). The scenarios also 
investigate different observing strategies. We do not explore all possible survey scenarios, but choose examples 
to highlight the flexibility of the modelling. The three ingredients in each scenario are as follows. 

\subsubsection{CTI Residual Modelling} 
Charge Transfer Inefficiency (CTI) is an effect caused in CCDs that are exposed to radiation. The radiation causes 
defects in the CCD in which electrons become trapped, which manifests itself as a blurred image in the readout direction of the 
CCD. Therefore the expected residuals should be limited to a CCD chip-scale and in the direction of the readout registers. 
To model the charge trailing due to CTI (see Massey et al., 2014), we use a model 
where a maximum residual ellipticity, or size, after correction is assigned at the center of a chip and linearly decreases to zero towards the chip 
edge. The amplitude of the maximum ellipticity, and the slope of the linear function can be changed on a chip-by-chip basis or 
over a field-of-view. The maximum ellipticity residual allowed due to 
imperfect corrective CTI 
modelling is taken from Cropper et al. (2013) to be $2.3\times 10^{-4}$, and the maximum fractional size residual is set to 
$5\times 10^{-4}$ to match the maximum PSF size (although this requirement is not set in Cropper et al., 2013). 

\subsubsection{PSF Residual Modelling}
To model residual ellipticity and size resulting from imperfect PSF modelling we use a polynomial of the form 
\ba
\delta \epsilon_{\rm PSF}&=&[c_0 + (c_1 x) + (c_2 y) \nn 
&+& (c_3 x^2) + (c_4 xy) + (c_5 y^2)\nn 
&+& (c_6 x^3) + (c_7 y^3)][1 + (c_8 x^2) + (c_9 y^2)]^{-1}
\ea
from Hoekstra (2004) that looked at the impact of PSF variation on shear correlation functions, where $x$ and $y$ are Cartesian coordinates in the field-of-view. The size 
residuals are assumed to have the same spatial behaviour as those of the ellipticity, although this assumption could be relaxed. The 
coefficients are chosen to create PSF patterns that have features in them that may be expected from data 
shown in Section \ref{Scenarios}. The polynomial variation is scaled such that the 
maximum ellipticity residual in the PSF residuals is $1.1\times 10^{-4}$ from Cropper et al. (2013), and the 
maximum fractional size residual is $5\times 10^{-4}$.  

\subsubsection{Shape Measurement Modelling}
In the scenarios we investigate we assume that half of the survey can be observed with four exposures and half with 
three exposures; this is a pessimistic implementation of a fiducial Euclid survey (e.g. Amiaux et al., 2012). Biases in shape measurement are dominated by uncertainty in the signal-to-noise of the 
galaxy images (see for example Viola, Kitching, Joachimi, 2014), here we assume that residual uncertanties 
are a function of the initial bias and hence  we simulate spatial variation in the multiplicative 
and additive parameters $\delta\mu$ and $\delta\alpha$ as being proportional to the number of exposures, that varies on field-of-view 
scales. The spatial variation is scaled such that the maximum values of these fields are $5\times 10^{-4}$ and $2\times 10^{-3}$ 
for $\alpha$ and $\delta\mu$ respectively (defined in Cropper et al., 2013). 

We emphasise that these models are only examples, and that in fact any functional or non-parameter spatial behaviour could have been 
included in this analysis. 

\subsection{Scenarios}
\label{Scenarios}
\begin{figure*}
\centering
  \includegraphics[angle=0,clip=,width=1.5\columnwidth]{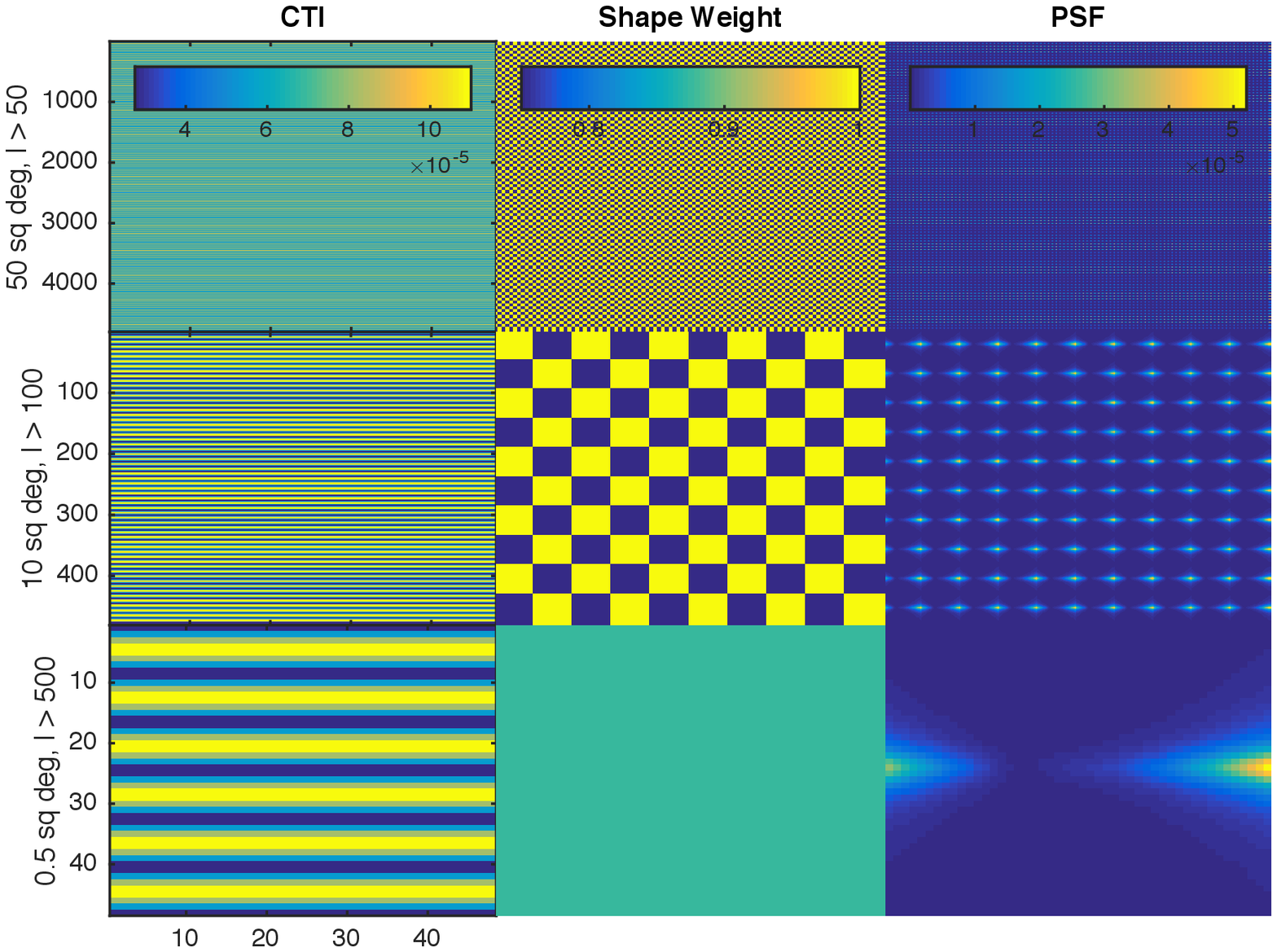}
  \includegraphics[angle=0,clip=,width=1\columnwidth]{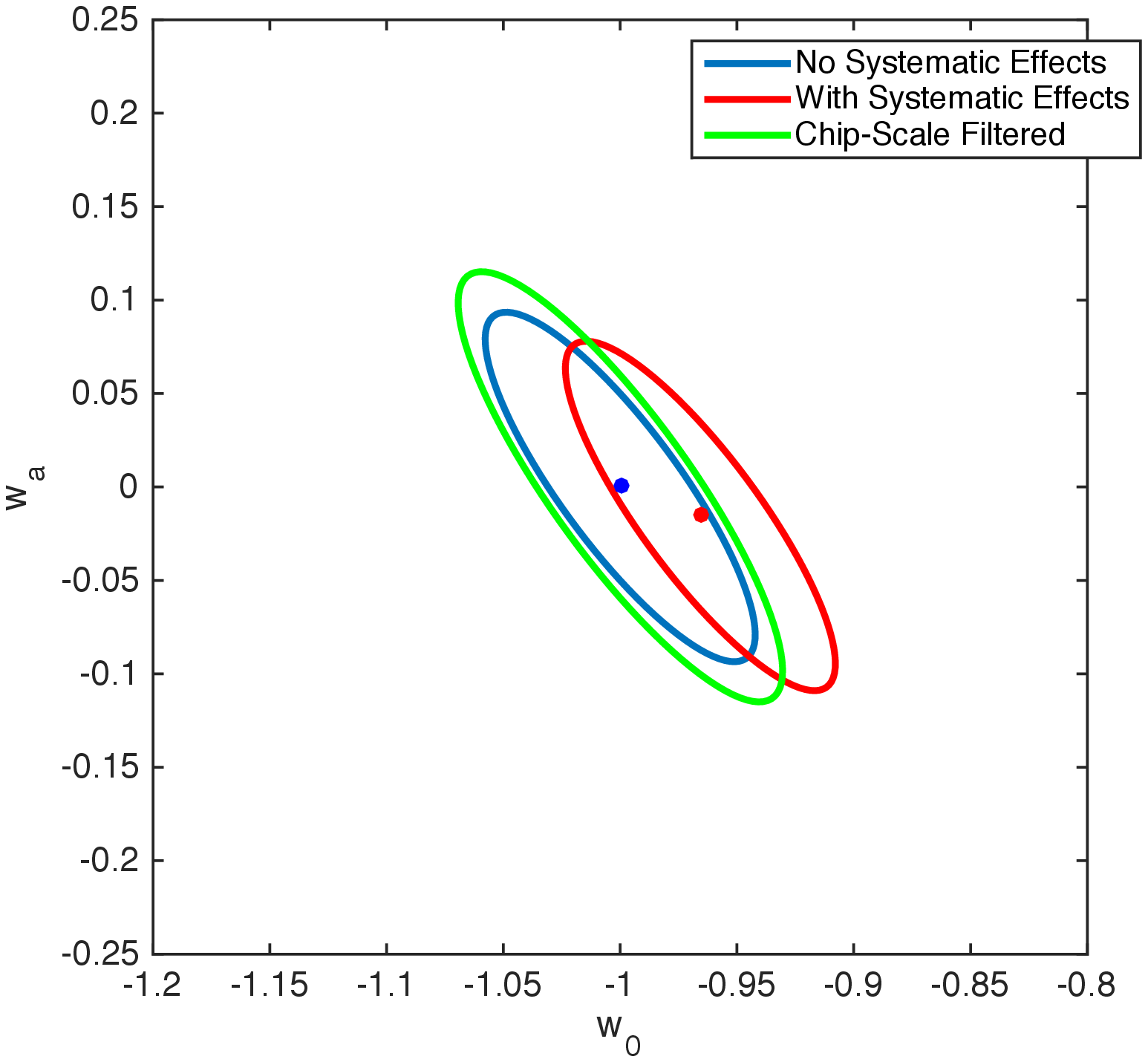}
  \includegraphics[angle=0,clip=,width=1\columnwidth]{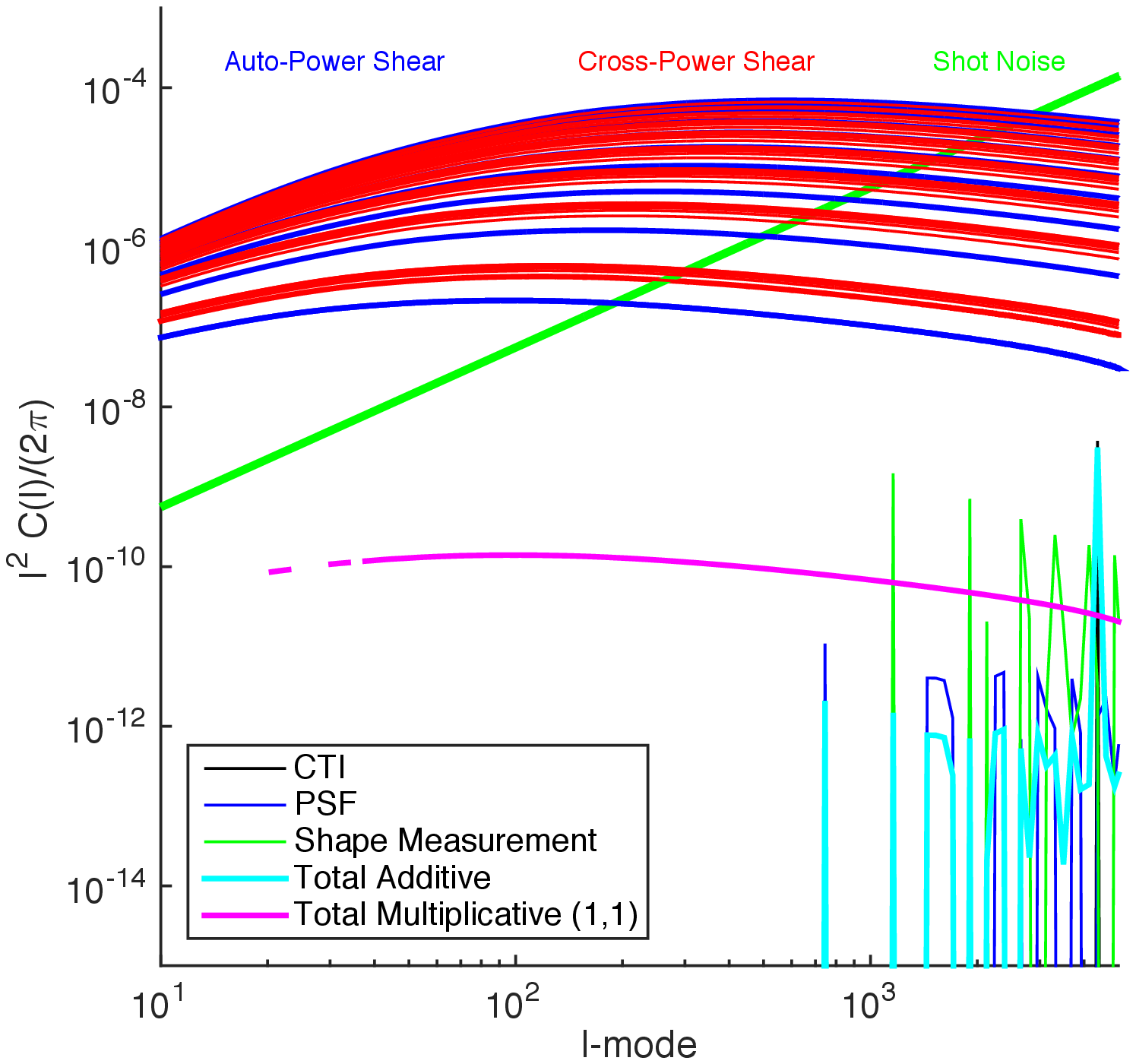}
  \includegraphics[angle=90,clip=,width=2\columnwidth]{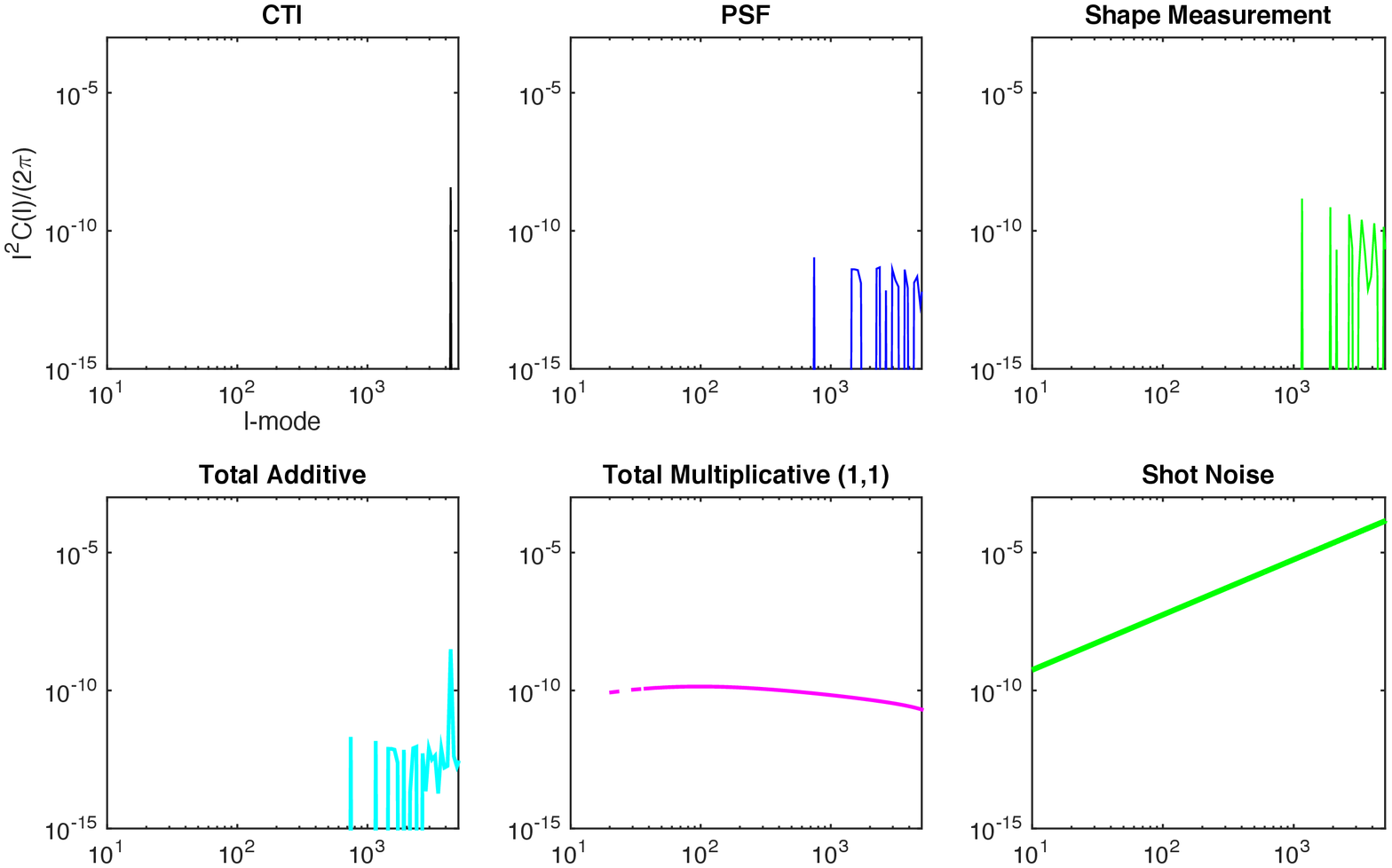}
 \caption{Scenario 1 systematic effects. The upper panels show the spatial variation in ellipticity for CTI, weighting and PSF (left to right), from
top to bottom these are progressive zooms-in over $50$ square degrees, $10$ square degrees, and
over $0.5$ square degrees (a field-of-view); the size variations have a similar pattern.
The middle right panel shows the power spectra for each ellipticity systematic effect, the cosmic shear tomography power spectra, and the per-$\ell$-mode
shot noise power spectrum.
The total systematic power spectra are shown (both the additive ${\mathcal A}_{\ell}$ and
multiplicative ${\mathcal M}_{\ell}$ terms) for the lowest redshift auto-correlation power spectra.
In the middle left panel we show the predicted marginalised 2-parameter 1-$\sigma$ error bars
in the $(w_0,w_a)$ plane with and without the systematic effects included for this case, and also in the case that the
chip-scale $\ell$-modes are filtered. The lower panels show the individual power spectra for each systematic effect.
The survey parameters are described in Section \ref{PM}.}
 \label{example1}
\end{figure*}

\begin{figure*}
\centering
  \includegraphics[angle=0,clip=,width=1.5\columnwidth]{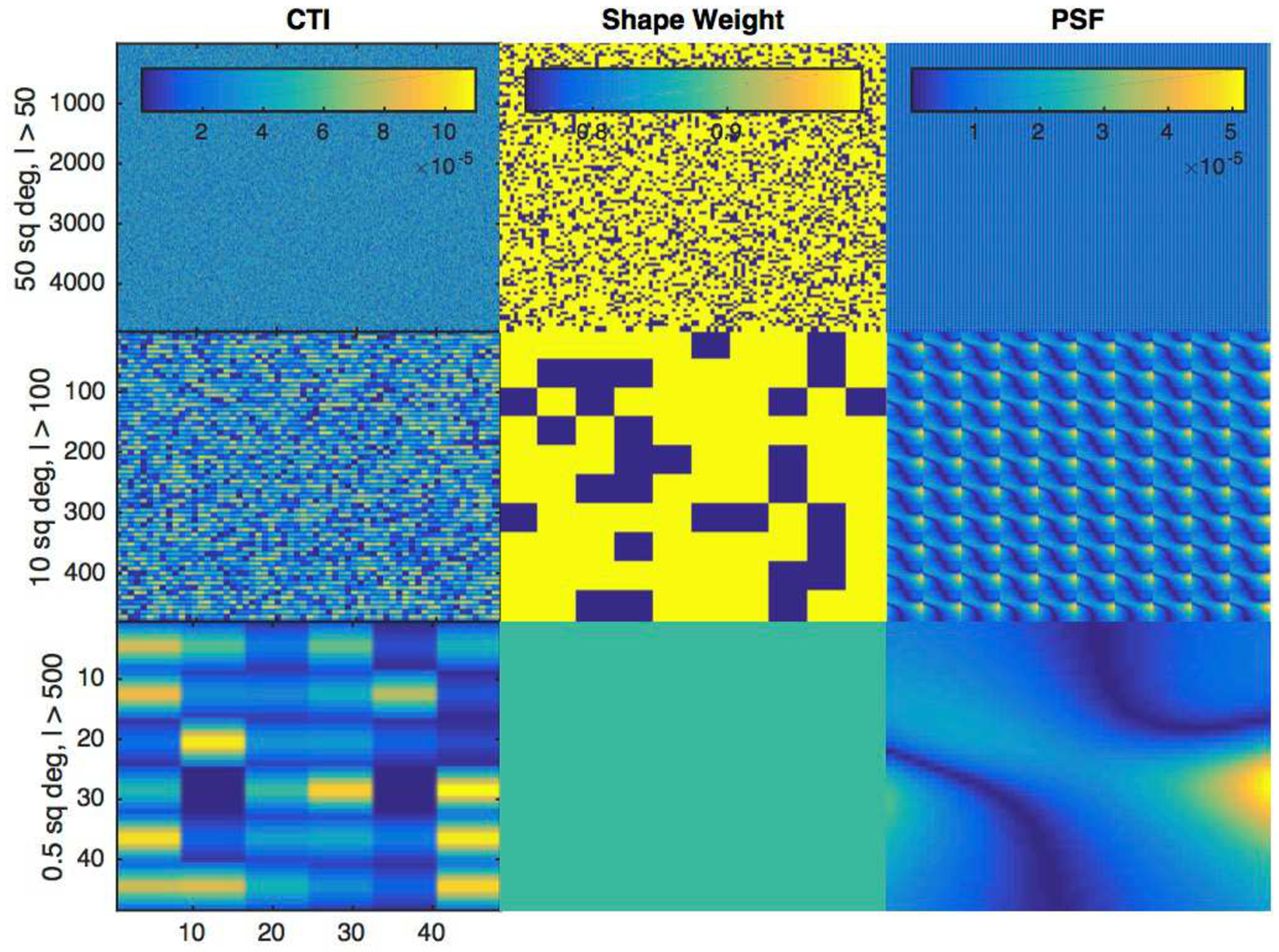}
  \includegraphics[angle=0,clip=,width=1\columnwidth]{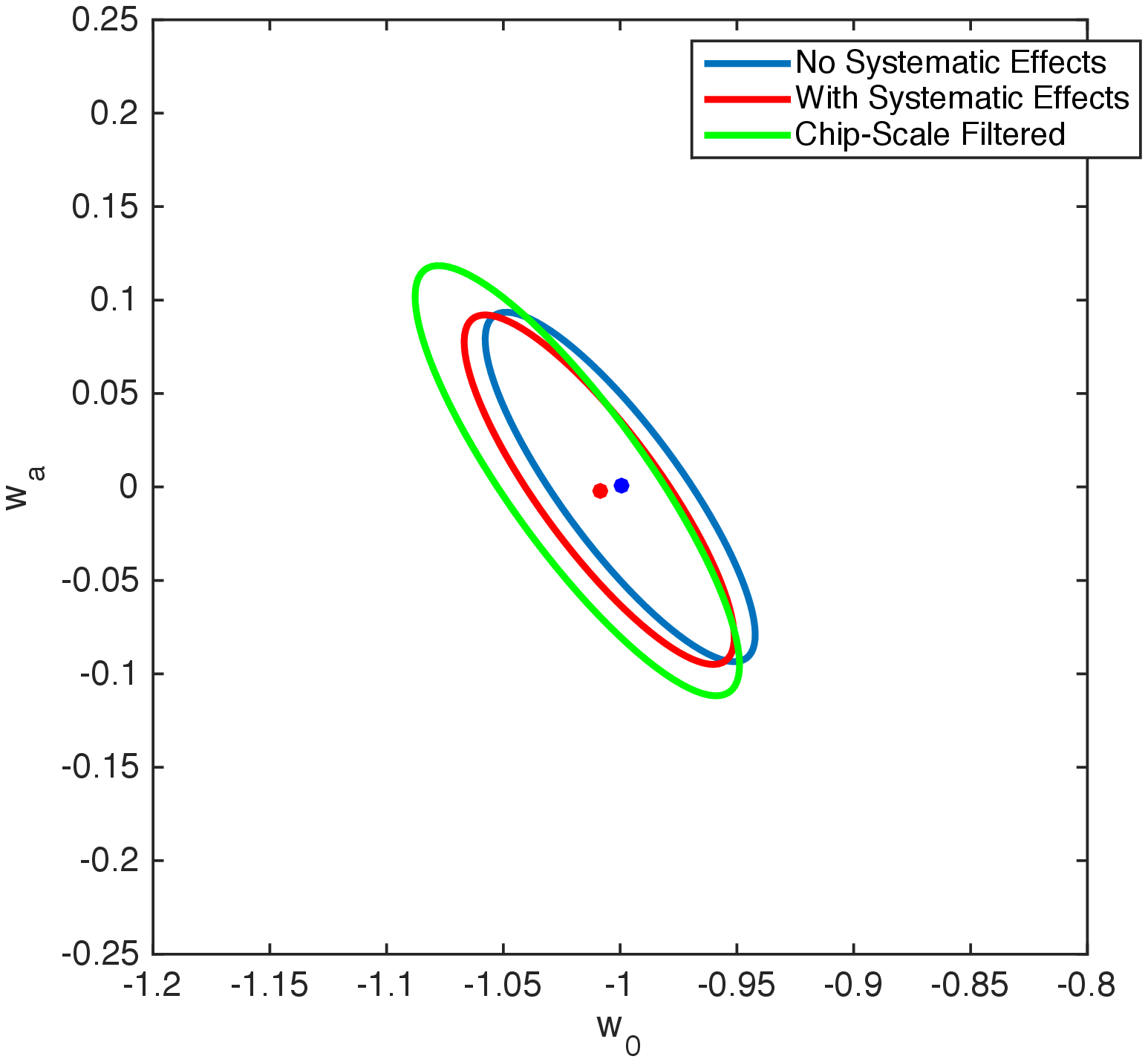}
  \includegraphics[angle=0,clip=,width=1\columnwidth]{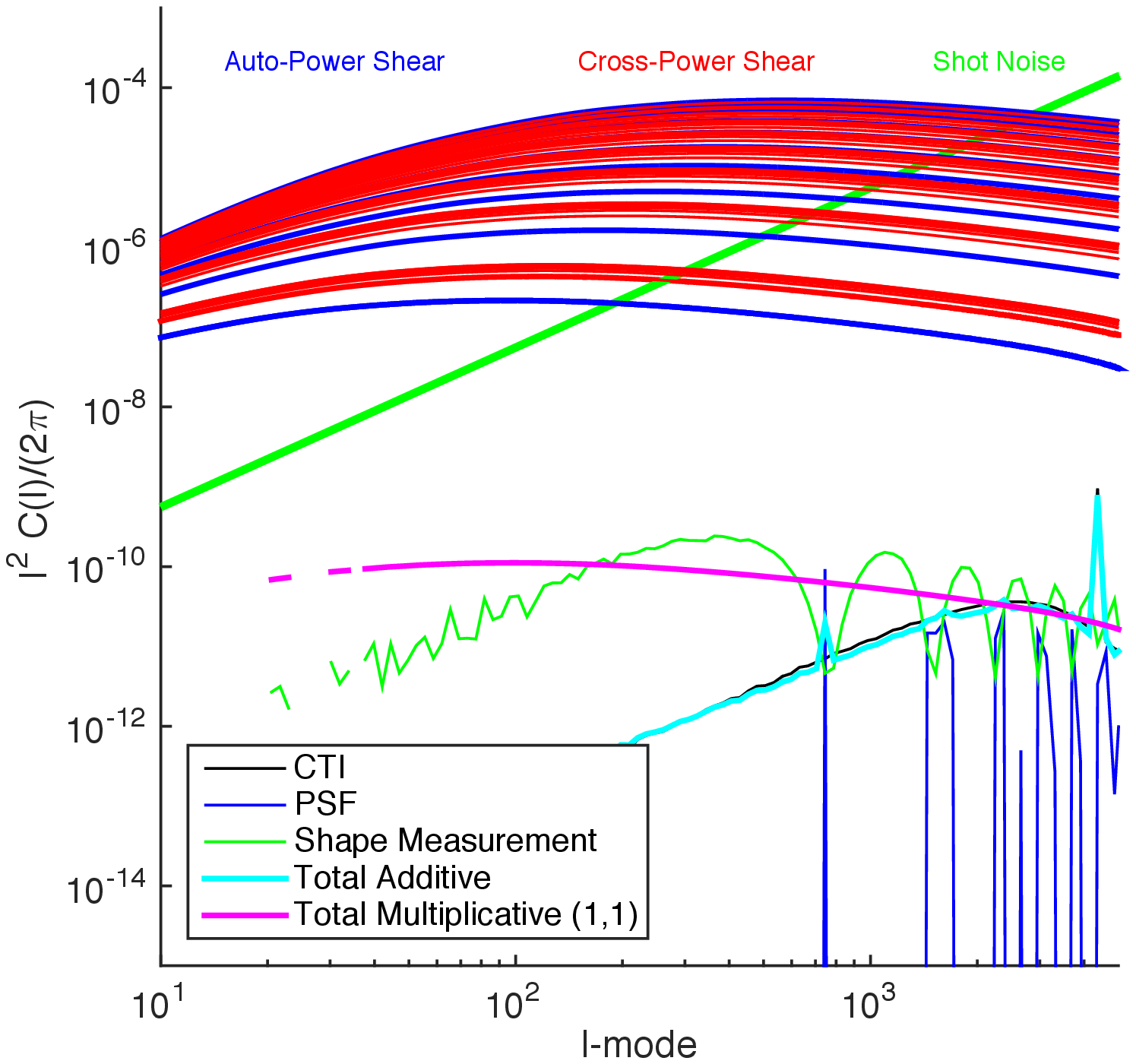}
  \includegraphics[angle=90,clip=,width=2\columnwidth]{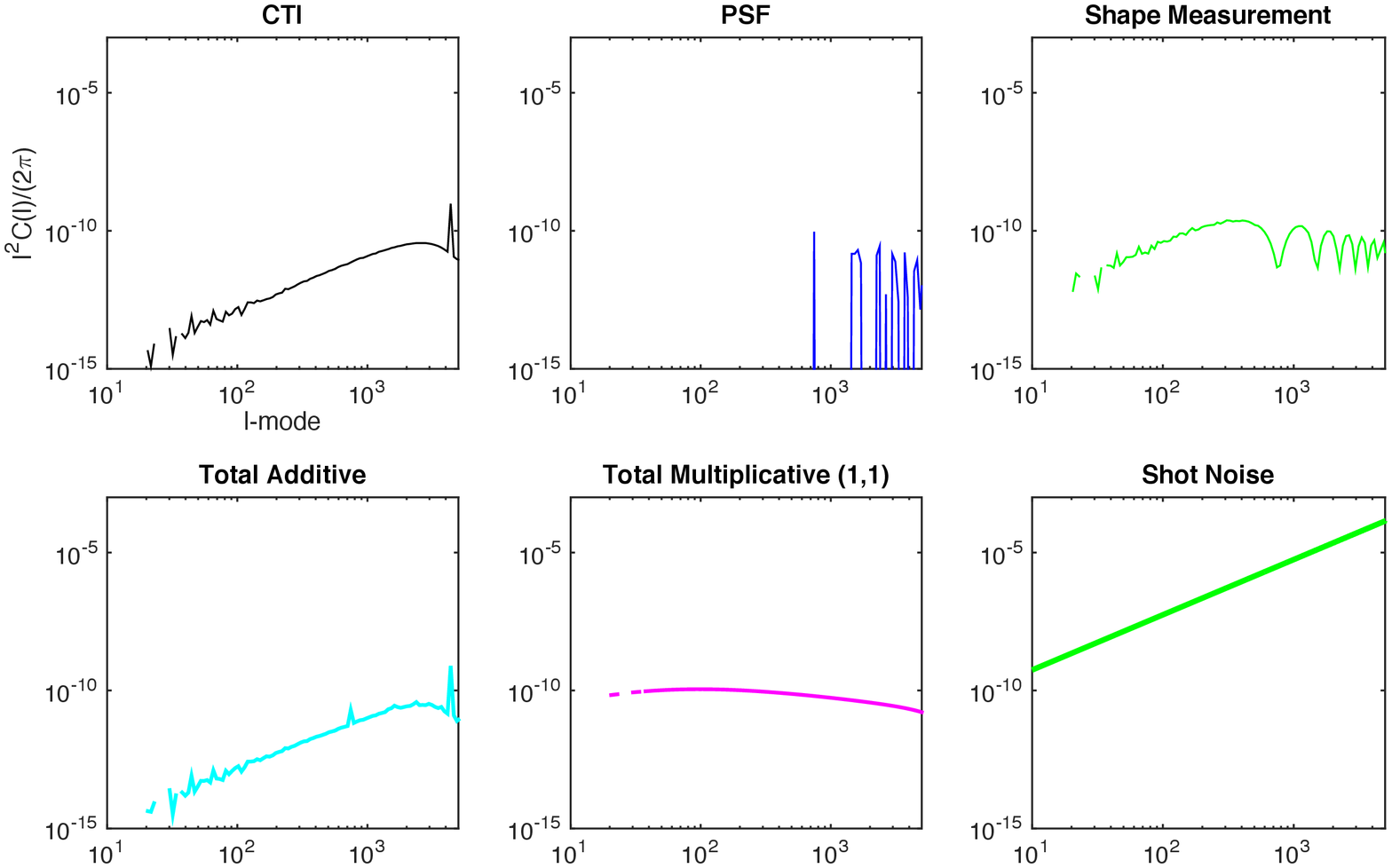}
 \caption{Scenario 2 systematic effects. The caption is the same as Figure \ref{example1}.}
 \label{example2}
\end{figure*}

\begin{figure*}
\centering
  \includegraphics[angle=0,clip=,width=1.5\columnwidth]{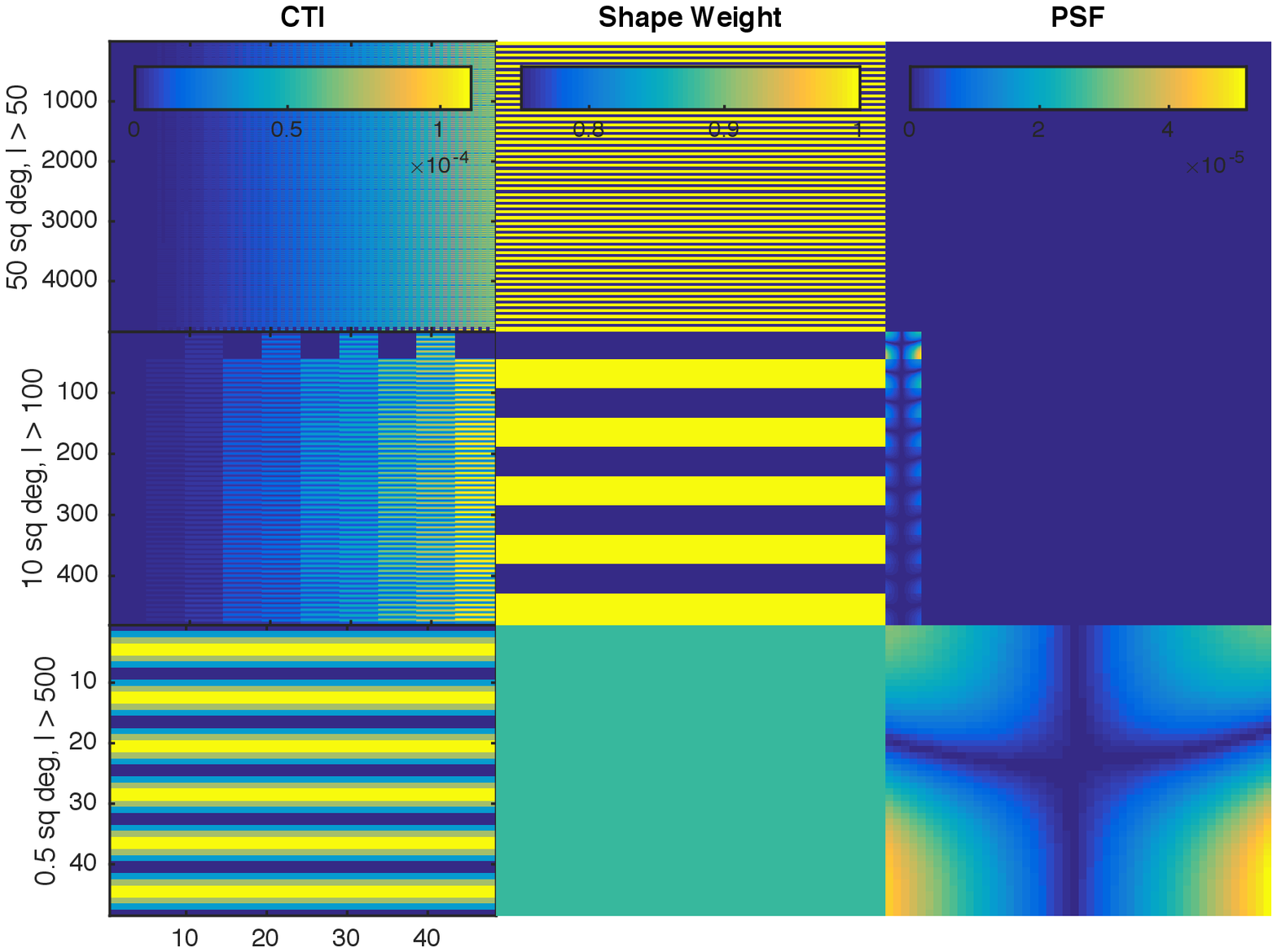}
  \includegraphics[angle=0,clip=,width=1\columnwidth]{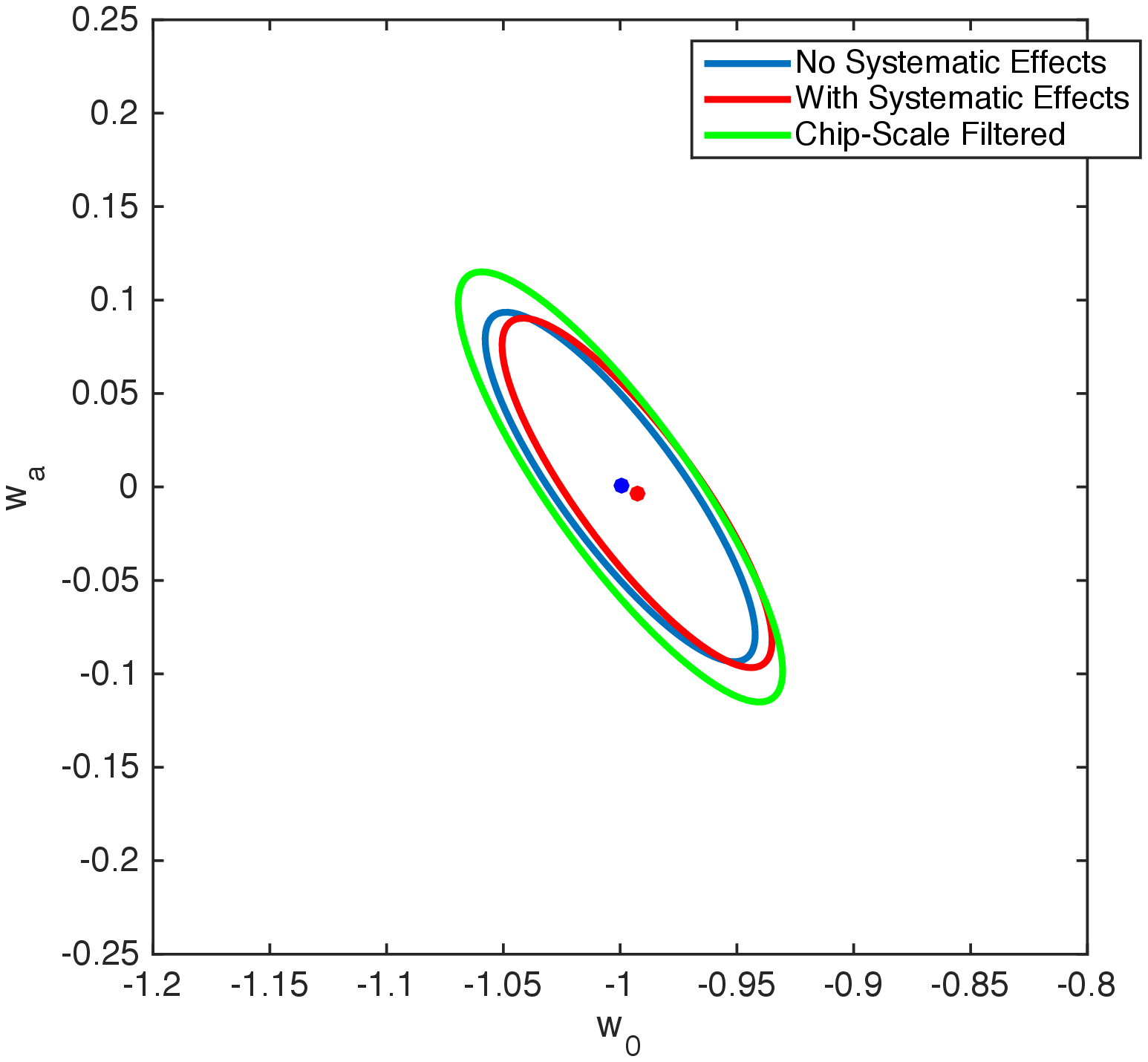}
  \includegraphics[angle=0,clip=,width=1\columnwidth]{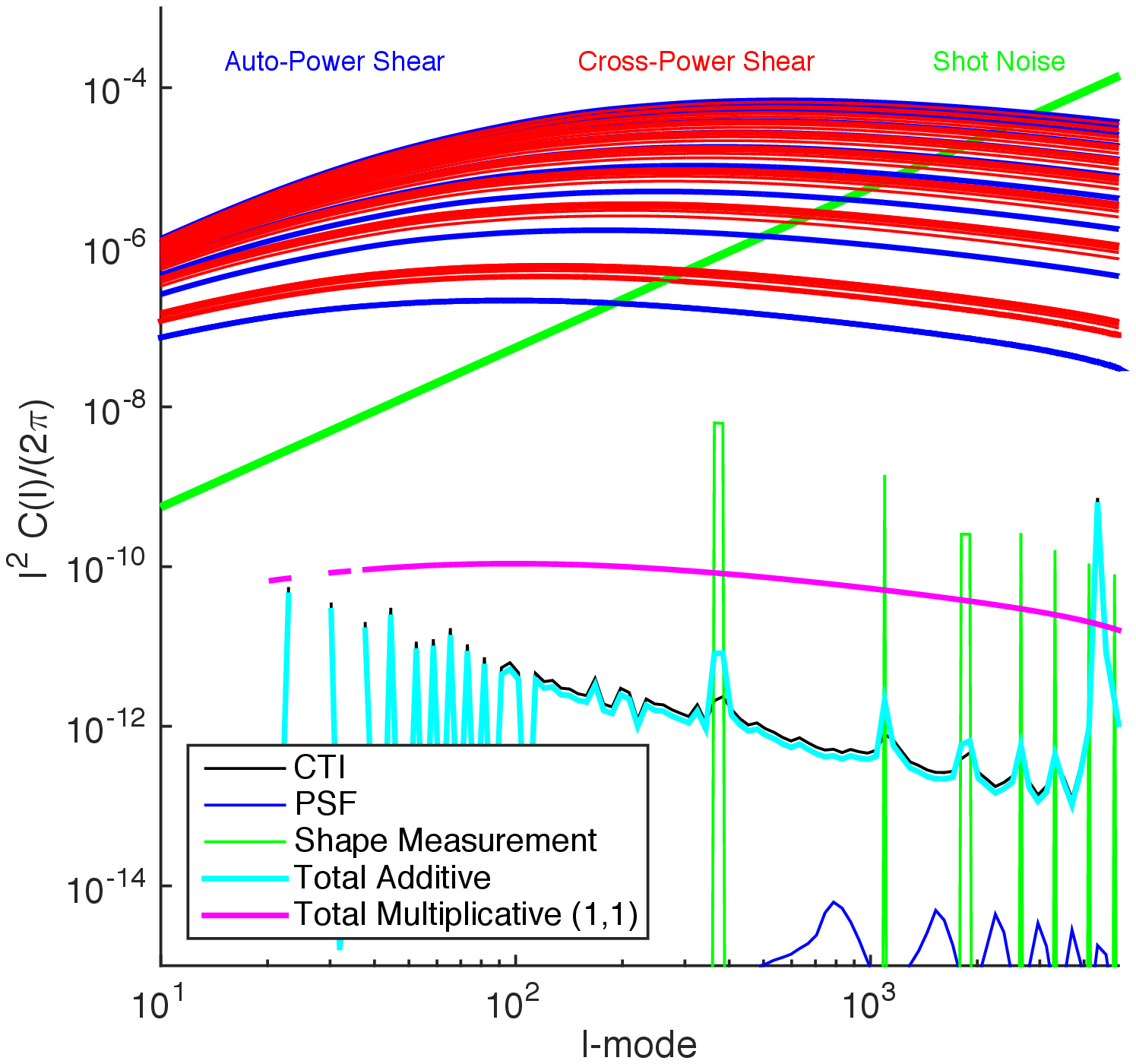}
  \includegraphics[angle=90,clip=,width=2\columnwidth]{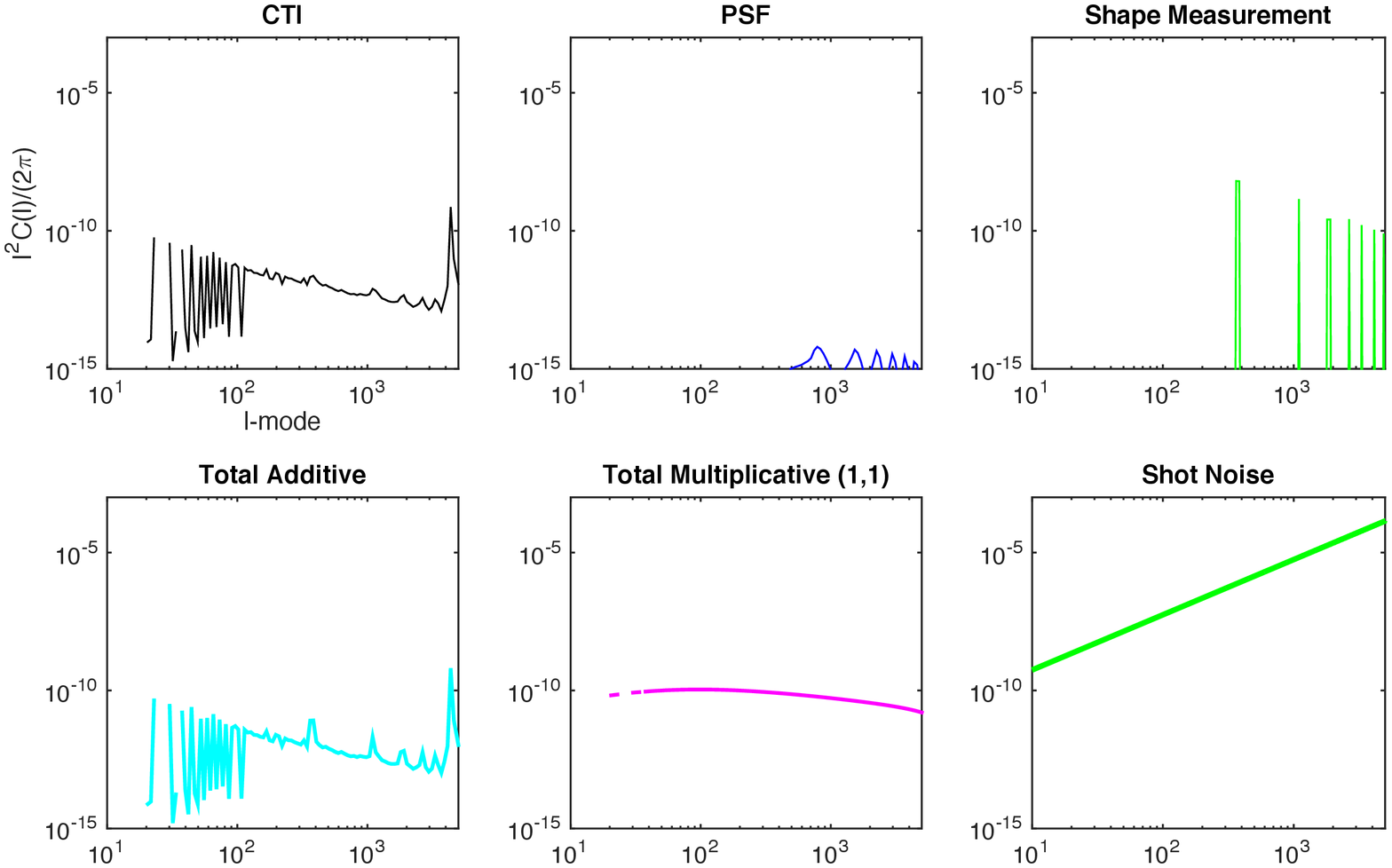}
 \caption{Scenario 3 systematic effects. The caption is the same as Figure \ref{example1}.}
 \label{example3}
\end{figure*}

Using the descriptions from the previous section we examine three Scenarios with variations about these basic prescriptions, 
described in Table \ref{scenes}. These scenarios were chosen to represent extreme cases in survey 
design and systematics modelling. In particular the way that systematics evolve over the survey were 
taken to be constant, random or evolving which are distinct categories. 
Figures \ref{example1}, \ref{example2} and \ref{example3} show the spatial patterns that the scenarios induce. 

In Scenario $1$ the CTI model and PSF model are constant for all fields-of-view: the values of the PSF coefficients used are 
$c_0 = 0.5$, $c_1 = 1.5$, $c_2 = 0.01$, $c_3 = 1.4$, $c_4 = 0.8$, $c_5 = 1$, $c_6 = 10$, $c_7 = 0.01$, $c_8 = 0.01$, $c_9 = 0.01$, 
the weighting function used is a checkerboard patter with alternative field-of-view having weights of $1$ and $0.75$. Scenario 2 
has a different residual pattern for every field of view, which is randomly assigned: the CTI amplitude is chosen to be 
between zero and the requirement stated in Section 2, the weight functon is also randomly assigned a value of either 
$1$ or $0.75$ for each field-of-view, the PSF has a wider residual pattern with coefficient values 
$c_0 = 100$, $c_1 = 2.5$, $c_2 = 0.1$, $c_3 = 2.4$, $c_4 = 1000$, $c_5 = 1$, $c_6 = 200$, $c_7 = 0.1$, $c_8 = 0.1$, $c_9 = 0.1$. 
Scenario 3 has the PSF and CTI residual patterns changing slowly over the survey area: the CTI amplitude linearly increases 
between each field-of-view, starting with zero and increasing in stripes over the survey, the PSF amplitude decreases between 
each field-of-view, starting with the requirement value and decreasig to zero, the PSF model has coefficient values
$c_0 = 0.5$, $c_1 = 2.5$, $c_2 = 0.1$, $c_3 = 2.4$, $c_4 = 100$, $c_5 = 1$, $c_6 = 20$, $c_7 = 0.1$, $c_8 = 0.1$, $c_9 = 0.1$, and 
the weight function has value of $1$ and $0.75$ arranged in stripes that are perpendicular to the direction over which the 
CTI and PSF change. These are realisations from a wide range of possible configurations that were chosen to be particulary  
extreme. 
\begin{table*}
\begin{tabular}{|l|c|l|l|l|}
&Scenario & {\bf 1) Constant Modelling} & {\bf 2) Randomised Model} & {\bf 3) FoV Evolution}\\
\hline
1.&CTI behaviour & Constant $\forall$ chip and FoV & Random amplitude & Increase in amplitude\\
2.&PSF behaviour & Constant \& compact variation &  Large random variation & Decrease in amplitude\\
3.&Field of View Tiling & Checkerboard & Random & Stripes\\
\hline
\end{tabular}
\caption{The three example scenarios. The constant modelling scenario has no field-to-field variation of either CTI or PSF, and 
a checkerboard pattern of shape measurement weighting. In the randomised case the CTI, PSF and weighting 
vary randomly across the survey area. 
In the evolution case we assume a rectilinear scanning strategy over the $50$ square degrees where the 
CTI gets progressively worse from the first to
last field-of-view, and the PSF model gets progressively better. The PSF polynomial is chosen to 
represent a range of cases over the three 
scenarios as used in Figures \ref{example1}, \ref{example2} and \ref{example3}.}
\label{scenes}
\end{table*}
We find that for all cases the dark energy Figure-of-Merit, $1/(\sigma(w_0)\sigma(w_a)-[\sigma^2(w_0,w_a)])$, is approximately $100$, similar to that found in Laureijs et al. (2011) for cosmic shear tomography alone with $10$ redshift bins centred on 
$[0.2, 0.4, 0.6, 0.8, 1.0, 1.2, 1.4, 1.6, 1.8, 2.0]$. 

Figures \ref{example1}, \ref{example2} and \ref{example3} show the power spectrum from each component of the introduced systematic effects. It can be seen that the repeated spatial pattern of the systematics - on chip and field-of-view scales 
($\ell_{\rm chip} \sim 2\pi/(0.5\pi/180/6) = 4320$)  - causes distinct features in the 
systematic power spectrum $C^{\rm sys}_{\ell}$ at these scales and their harmonics. 
The exact functional form depends on the distribution and the amplitude of the 
systematic effect modelled. Importantly in none of the extreme cases we consider does the systematic power spectrum mimic the 
derivative of the shear power spectrum with respect to any cosmological parameter. 
Hence in all three cases there is a small change in the predicted cosmological parameter errors, and bias, despite there being 
significant systematic effects in the modelled data. We find that, in agreement with the simple models in Section \ref{simp}, systematic effects 
with an approximately shot-noise like power spectrum have a smaller bias. We also try filtering these scales, 
as we describe in Section 4.3. 

The systematic effects that we simulate in this paper are all
redshift independent. Therefore the primary impact they have is on $w_0$ which is a constant with redshift; $w_a$ should be more affected by redshift-dependent 
systematic effects, and this is indeed seen in investigations on intrinsic alignments (see e.g. Kirk et al., 2015).

In Figure \ref{example1} the constant patterns produce variation only on the chip and field-of-view scales, hence the power spectra of the systematic 
effects are all at high-$\ell$, and in the case of CTI localised at the chip scale, or multiples thereof. Therefore the removal of the chip scale power is particularly effective at 
removing bias. 
In Figure \ref{example2} the randomised nature of the systematic effects spreads the power over all scales. In particular it can be seen that the 
single spike in power due to CTI now has a broad shot noise-like component. These systematic power spectra therefore cause smaller biases, but the 
less clearly defined spike at the chip scale means that the removal of these modes is less effective. The removal of modes also increases the error bar, through the loss of 
information. In Figure \ref{example3} the regular field-of-view scale variation 
adds power at these scales, and the gradual changes over the entire field add power at low-$\ell$. 

The values of the integrated additive and multiplicative contributions in each of the scenarios are for Figures \ref{example1}, \ref{example2}, and \ref{example3} : 
$\overline{{\mathcal A}}= 3.0\times 10^{-6}$ and $\overline{{\mathcal M}}=1.0\times 10^{-1}$,  
$\overline{{\mathcal A}}= 2.9\times 10^{-5}$ and $\overline{{\mathcal M}}=8.1\times 10^{-2}$, and 
$\overline{{\mathcal A}}= 3.4\times 10^{-5}$ and $\overline{{\mathcal M}}=7.6\times 10^{-2}$ respectively. These in some cases exceed 
the requirement set in Cropper et al. (2013) and Massey et al. (2013) despite the fact that the dark energy measurement would 
remain largely unaffected. This suggests that an improved set of requirements can be set, taking into account the expected 
spatial variation in the systematic effects. However we 
note that this can only be done once a model set up is evaluated for a particular experiment.  
 
\subsection{CTI Requirements}
\label{Requirements}
In each of the Figures \ref{example1}, \ref{example2} and \ref{example3} we show the effect of removing power that is present around the 
CCD scale, and find that the dark energy biases are reduced in doing this. We now 
explore this $\ell$-mode-filtering further within the context of the CTI systematic effect. 

What the examples in Sections \ref{simp} and \ref{Scenarios} show is that the dominant effect that causes a systematic bias in
cosmic shear is the regularity (or stochasticity), of the spatially varying nature of the effect, that focuses power at particular 
frequency ranges in the power spectrum. To explore this further
we consider requirements on one aspect of cosmic shear survey design: the amplitude of the detector CTI ellipticity residuals.

We set requirements on CTI by assuming a scenario where all the contribution to the total systematic additive bias comes from
CTI only. These requirements will set an upper limit on the contribution of CTI to the total error budget on cosmic shear
calculations. We assume that the CTI residual amplitude increases linearly in amplitude, 
similar to the ``field-of-view evolution'' 
scenarios in the previous
section, and we vary the size of the patch over which the rectilinear scanning strategy is constant. We look at two cases, one where the full
$50$ square degrees is scanned in a rectilinear fashion, and one where the scanning strategy is entirely random. 
In each case the spatial arrangement of the fields is the same, so that the maxima and minima of the CTI pattern are the same, and it is 
only the order with which the fields are observed, and hence the amplitude of the CTI (which grows with time) that is different at different parts in the field.
This is similar to the simple example given in Figure \ref{test2}.

\begin{figure*}
  \includegraphics[angle=0,clip=,width=\columnwidth]{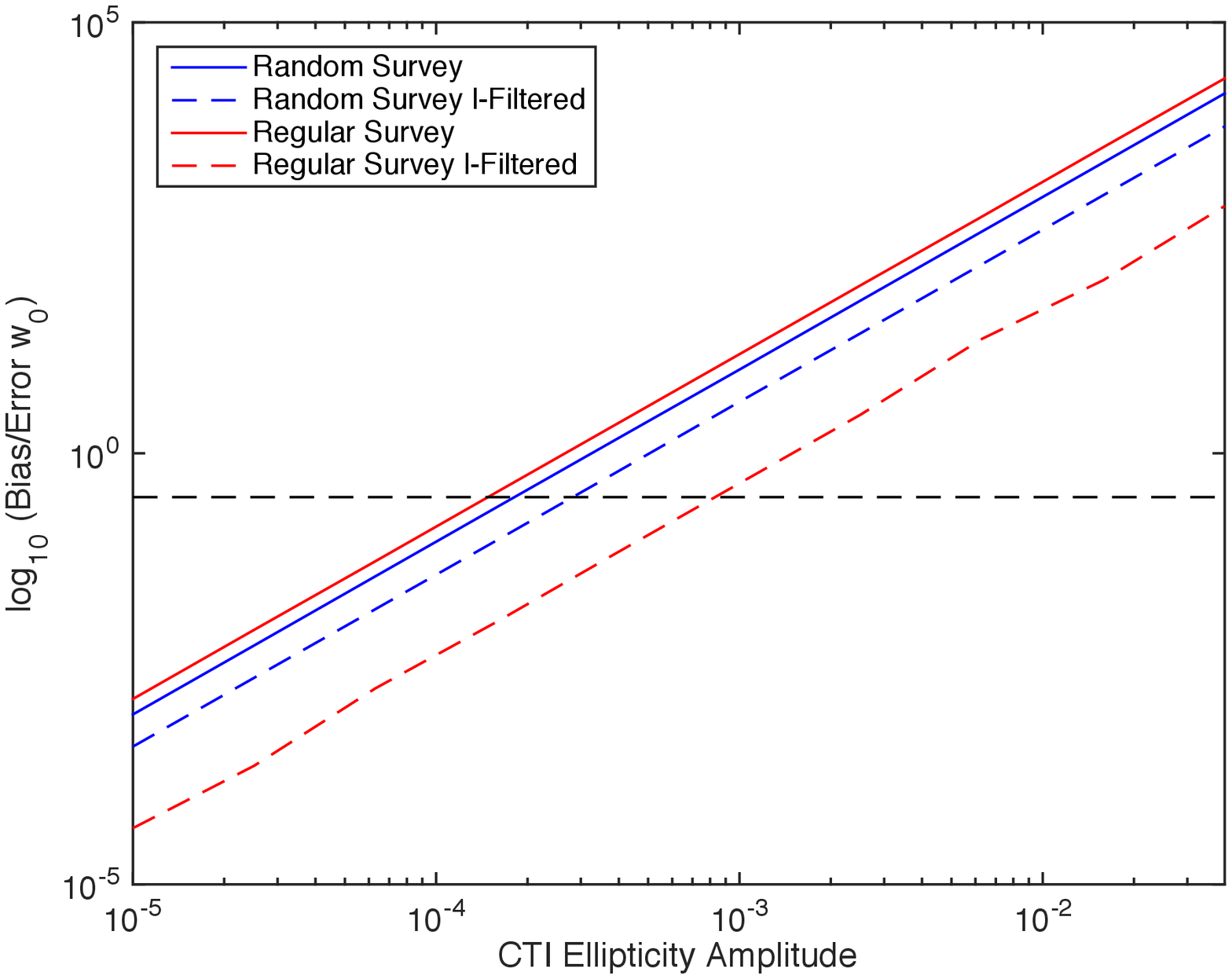}
 \caption{The change in dark energy parameter $w_0$ bias/error as a function of CTI ellipticity amplitude. Shown are solid lines 
with no $\ell$-mode filtering and dashed lines with $\ell$-mode filtering, for surveys that have a random surveying strategy (blue 
lines) and a rectilinear observing strategy (red lines). The horizontal dashed line shows the requirements from Massey et al. (2014) 
that the bias/error is less than $0.31$. The survey parameters are described in Section \ref{PM}.} 
 \label{ctireq}
\end{figure*}

In Figure \ref{ctireq} we show how the bias on the dark energy equation of state varies with CTI amplitude for regular and randomised scanning 
strategies. We find, consistent with the previous results in this paper, that a randomised scanning strategy has a lower bias than a
regular one. We also find that the requirement of $2.3\times 10^{-4}$ used in Cropper et al. (2013) is indeed a good level
for this systematic effect. However, given that the CTI effect creates a sharply peaked systematic effect at the scale of the detector
$\ell_{\rm chip}\approx 2\pi/(0.5\pi/180/6)=4320$ it should be possible to simply remove scales around this peak from the analysis to recover an
unbiased estimate of cosmological parameters. We test this $\ell$-mode-filtering by removing scales $\ell_{\rm chip}\pm 100$
from analysis. The dark energy Figure of Merit is reduced through the loss of information on these scales. However,  
this is still conservative as we could have allowed a coherent pattern in the detector frame that we then modelled instead of 
removed. In the absence of a model, the removal of these scales leads to a relatively modest reduction in Figure of Merit of $10\%$. 
In Figure \ref{ctireq} we show the impact of this filtering on the cosmology bias estimates. We find
that for the case of a single $50$ square degree patch the requirement on the CTI ellipticity amplitude is reduced by a factor of $9$, 
In this
case (as also shown in Figure \ref{test2}) the power spectrum is sharply peaked at the chip-scale. For the randomised patch design the
improvement is less, because the power is spread over a larger range of scales. Therefore we find that although a regularised survey
design has a slightly larger bias, compared to a randomised pattern, the removal of scales affected by systematics can reduce biases. This however will only work for additive-type biases as the multiplicative bias acts as a convolution in 
Fourier space. 

\begin{figure*}
\centering
 \includegraphics[angle=0,clip=,width=2\columnwidth]{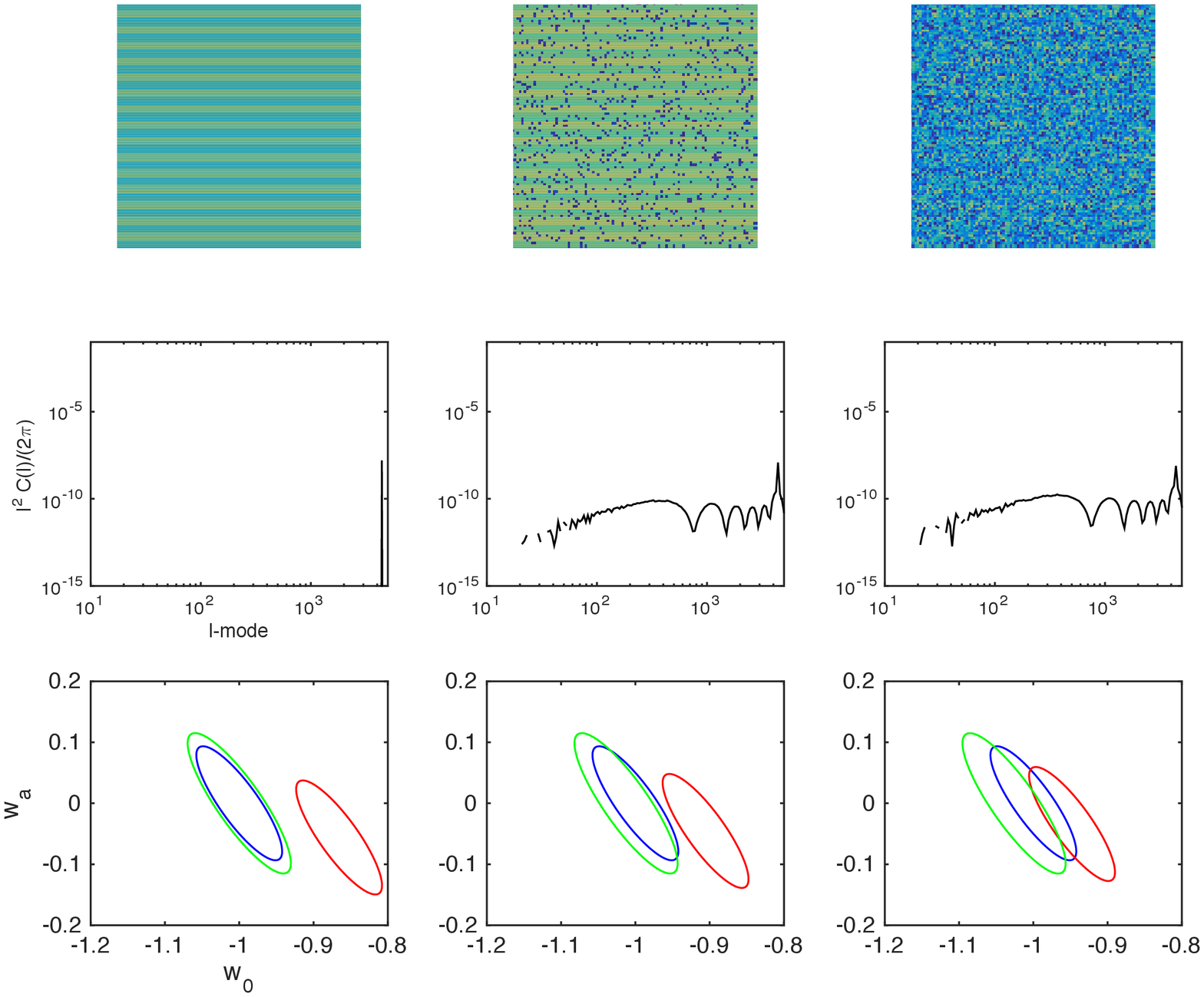}                                                           
 \caption{The top row of panels show three different cases where CTI amplitude changes over a $50$ square degree patch, 
where the colour represents residual ellipticity 
amplitude caused by CTI (red high and blue small): the left hand panel where the CTI is the same over the whole patch; the 
middle panel where CTI is the same for $90\%$ of the patch, and zero for the remainder; the right panel where the CTI amplitude 
randomly changes from field to field. The second row shows the corresponding power spectra for each systematic ellipticity field  
(the y-axes are $\ell^2 C(\ell)/(2\pi)$ and the x-axes are $\ell$-mode). The lower panels show the expected errors in the 
$(w_0,w_a)$ plane: the blue contours are with no CTI, the red contours with CTI, and the green contours with CTI and 
$\ell$-mode filtering. The survey parameters are described in Section \ref{PM}. } 
 \label{ctireq2}
\end{figure*}

We explore the concept of $\ell$-mode filtering further in Figure \ref{ctireq2} where we show three examples where the CTI is 1) constant
over the entire $50$ square degrees -- leading to a spike in the power spectrum at the chip scale; 2) evolves to reach a
maximum after over $10\%$ of fields are observed; 3) evolves to a maximum over the whole area. In this case we see that if the CTI
is constant, then it is easily removed and leads to negligible parameter biases. In each plot the harmonics of the chip scale are outside the plotting and 
maximum $\ell$-mode used. In the case that the CTI evolves, the power is spread
from the chip-scale to the field-of-view scale, meaning it is more difficult to remove the contaminated scales, causing larger parameter
biases. The difference in the power spectra between the second and third cases is very slight, an overall change in amplitude over most scales. 
                                   
\section{Conclusion}
\label{Conclusion}
In this paper we investigated how spatially varying systematic effects can propagate into tomographic cosmic shear power spectra, 
and into the biases on the dark energy equation of state parameters $w_0$ and $w_a$ inferred from these power spectra. 
This is a generalisation of previous studies that assumed systematics with no spatial variation 
(that affect all parts in a field-of-view equally), or random spatial variation. 
Many realistic systematics can be identified and nulled, and many residuals do not cause changes in the power spectrum 
that cause large biases in dark energy equation of state measurements. 
Consequently, both the previous approaches lead to requirements on weak lensing that are too stringent. 
Given that several surveys have been designed with these conservative assumptions in mind, this means 
that the expected performance is 
better or requirements can be relaxed (i.e. there is likely to be `margin' in the design).  

We find that the metric used to set requirements on systematic effects by e.g. Amara \& Refregier (2008) 
and Massey et al. (2013) -- namely the multiplicative and additive biases on the power spectrum integrated 
equally over all scales -- do not account for the full story. 
These requirements 
can be exceeded in many realistic scenarios whilst the dark energy measurement can remain unbiased. This all assumes that a survey is driven by 
dark energy measurements, but a further aim may be measure power spectra for more general purposes. An additional point is that most 
of the biases we find are present on small scales ($\ell > 1000$) which are generally more difficult to model astrophysically (due to 
example baryonic feedback effects). So a cleaner cosmological probe both in terms of systematic effects and poor modelling may be one that 
removes such scale from an analysis and supplements the weak lensing with addition data (e.g. information from clusters). 

Through a series of simplified examples we show 
that there is a trade-off between the spatial regularity of a systematic effect, its amplitude, and the integrated systematic power spectrum. 
A systematic effect that has a purely random spatial pattern acts to introduce shot-noise-like power to the cosmic shear, and as a result only 
slightly increases the cosmological parameter uncertainties -- without causing a bias in the parameters. Conversely a very regular spatial pattern will 
cause sharply peaked features in the power spectrum, that can cause large biases but not impact the error bar. However if a systematic effect causes localised changes 
in the power spectrum, these are easy to remove by ignoring those $\ell$-modes in a cosmological analysis; this is a conservative approach 
because one could also model these changes in the power spectrum.

We show three example scenarios that include modelling of CCD CTI effects, shape measurement error variation, and PSF modelling variation. These 
are extreme examples that serve to highlight that even in these cases the effect on dark energy equation of state measurements is small.  
We find that a significant contributor to bias in the dark energy equation of state comes from the survey observing pattern. 
If a survey has a 
randomised pattern of field-to-field weighting this acts to spread the systematic power over a larger range of scales. In contrast a 
rectilinear or striped pattern concentrates power on smaller scales that can be more easily filtered. 

It should be noted however that 
throughout we make a number of simplifying assumptions, for example we do not include the effects of image `dithering', and use a flat-sky approximation. The image dithering, and the use of multiple $50$ square degree patches over the sky in the full survey, the phasing of the effects in each of which will not necessarily be maintained, will also qualify the effectiveness of the filtering. 
In particular a sharp spike-like feature on the chip-scale may be spread out over several $\ell$-modes in a similar way to that seen 
in the randomised scenerio we investigate. In this case the filtering will become less effective, but the systematic itself will be closer to a randomised patterm which is likely to reduce bias in any case. We leave a full investigation of dithering effects for future work. 

We use the formalism to investigate requirements on the amplitude of the CTI effect for cosmic shear. We find that in the case that  
the survey strategy is random, the contribution to the power spectrum is spread over a wide range in $\ell$-mode and that the result from previous studies are recovered: that the 
ellipticity amplitude should be less than $2\times 10^{-4}$. If the survey strategy is more regular then the systematic power spectrum is concentrated at a particular 
scale -- with a larger amplitude -- and causes larger biases. However because the systematic effect is very localised in $\ell$-mode then a straightforward 
mitigation strategy is to remove those scales from a cosmological analyses and we find that such $\ell$-mode-filtering 
reduces cosmological parameter biases relative to the randomised survey strategy case at the 
expenses of a $10\%$ increase in the size of error bars.

The framework presented in this paper is a generalisation of investigations into systematic effects in cosmic shear surveys. Such a framework can be used to 
build more optimal survey strategies, and find margin in imaging survey specifications, potentially leading to less stringent requirements on the control of 
systematic effects in instrument and algorithmic design.


\noindent{\em Acknowledgements:} TDK and RM are supported by Royal Society University Research Fellowships. RM acknowledges the Science and Technology Facilities Council (grant number ST/H005234/1) and the Leverhulme Trust (grant number PLP-2011-003).
Part of this work was an extension of the work 
presented in Hood (2014), a UCL Space Science and Engineering MSc thesis project (supervisor TDK). 



\onecolumn


\begin{thebibliography}{}

\bibitem[Amara \& R{\'e}fr{\'e}gier(2008)]{2008MNRAS.391..228A} Amara, A., \& R{\'e}fr{\'e}gier, A.\ 2008, \mnras, 391, 228 
\bibitem[Amiaux et al.(2012)]{2012SPIE.8442E..0ZA} Amiaux, J., Scaramella, R., Mellier, Y., et al.\ 2012, Proc. SPIE, 8442, 84420Z 
\bibitem[Bartelmann \& Schneider(2001)]{2001PhR...340..291B} Bartelmann, M., \& Schneider, P.\ 2001, physrep, 340, 291 
\bibitem[Battye et al.(2014)]{2014arXiv1409.2769B} Battye, R.~A., Charnock, T., \& Moss, A.\ 2014, arXiv:1409.2769
\bibitem[Cropper et al.(2013)]{2013MNRAS.431.3103C} Cropper, M., Hoekstra, H., Kitching, T., et al.\ 2013, \mnras, 431, 3103 
\bibitem[Hoekstra \& Jain(2008)]{2008ARNPS..58...99H} Hoekstra, H., \& Jain, B.\ 2008, Annual Review of Nuclear and Particle Science, 58, 99 
\bibitem[Hu(1999)]{1999ApJ...522L..21H} Hu, W.\ 1999, apjl, 522, L21 
\bibitem[Hood (2014)]{}Hood, R., UCL MSc Thesis, UCL; supervisor T. D. Kitching
\bibitem[Joachimi et al.(2015)]{2015arXiv150405456J} Joachimi, B., Cacciato, M., Kitching, T.~D., et al.\ 2015, arXiv:1504.05456 
\bibitem[Kilbinger(2014)]{2014arXiv1411.0115K} Kilbinger, M.\ 2014, arXiv:1411.0115 
\bibitem[Kirk et al.(2015)]{2015arXiv150405465K} Kirk, D., et al.\ 2015, arXiv:1504.05465 
\bibitem[Kitching et al.(2009)]{2009MNRAS.399.2107K} Kitching, T.~D., Amara, A., Abdalla, F.~B., Joachimi, B., \& Refregier, A.\ 2009, \mnras, 399, 2107 
\bibitem[Kitching et al.(2011)]{2011MNRAS.413.2923K} Kitching, T.~D., Heavens, A.~F., \& Miller, L.\ 2011, \mnras, 413, 2923 
\bibitem[Kitching et al.(2012)]{2012MNRAS.423.3163K} Kitching, T.~D., Balan, S.~T., Bridle, S., et al.\ 2012, \mnras, 423, 3163 
\bibitem[Kitching \& Taylor(2011)]{2011MNRAS.416.1717K} Kitching, T.~D., \& Taylor, A.~N.\ 2011, \mnras, 416, 1717
\bibitem[Laureijs et al.(2011)]{2011arXiv1110.3193L} Laureijs, R., et al.\ 2011, arXiv:1110.3193 
\bibitem[MacCrann et al.(2014)]{2014arXiv1408.4742M} MacCrann, N., Zuntz, J., Bridle, S., Jain, B., \& Becker, M.~R.\ 2014, arXiv:1408.4742 
\bibitem[Massey et al.(2013)]{2013MNRAS.429..661M} Massey, R., Hoekstra, H., Kitching, T., et al.\ 2013, \mnras, 429, 661 
\bibitem[Massey et al.(2014)]{2014MNRAS.439..887M} Massey, R., Schrabback, T., Cordes, O., et al.\ 2014, \mnras, 439, 887 
\bibitem[Troxel \& Ishak(2014)]{2014arXiv1407.6990T} Troxel, M.~A., \& Ishak, M.\ 2014, arXiv:1407.6990 
\bibitem[Viola et al.(2014)]{2014MNRAS.439.1909V} Viola, M., Kitching, T.~D., \& Joachimi, B.\ 2014, \mnras, 439, 1909 
\bibitem[Voigt \& Bridle(2010)]{2010MNRAS.404..458V} Voigt, L.~M., \& Bridle, S.~L.\ 2010, \mnras, 404, 458

\end{thebibliography}
\end{document}